\documentclass[12pt,a4paper]{article}
\usepackage[english]{babel}
\usepackage[T1]{fontenc}
\usepackage[adobe-utopia]{mathdesign}
\usepackage[intlimits]{amsmath}
\usepackage{graphicx}
\usepackage{a4wide}
\usepackage{amsbsy}
\usepackage{hyperref}
\usepackage{physics}
\usepackage[square, authoryear]{natbib}
\usepackage{authblk}
\bibpunct{[}{]}{;}{a}{}{,}
\DeclareMathSymbol{\mlq}{\mathord}{operators}{``}
\DeclareMathSymbol{\mrq}{\mathord}{operators}{`'}

\begin{document}

\title{\textsc{\huge{Tokyo Wheeler \emph{or} the Epistemic Preconditions of the Renaissance of Relativity}}}
\author[1]{Alexander Blum}
\author[2]{Dieter Brill}
\affil[1]{Max Planck Institute for the History of Science, Berlin, Germany}
\affil[2]{Department of Physics, University of Maryland, College Park, MD, USA}
\maketitle

\begin{abstract}

During the period identified as the Rebirth of General Relativity, John Wheeler was instrumental in retrieving the physics of gravity that had become hidden behind the mathematical formalism. For Wheeler himself the change in point of view was not from mathematics to physics; his thinking about gravity arose from, and was largely guided by, physical problems about elementary particles. We recount his development from the view of fields as derivable from particles, to fields as the more fundamental entities in nature. During the more than 10 years of his search for "particles first" Wheeler did not write an orderly sequence of accounts as his views developed, but the story could be pieced together from letters. and particularly from his extensive but seldom sequential notebook entries.

Our story starts with Wheeler and Feynman's reformulation of classical electrodynamics that allows an elimination of the field degrees of freedom and replacement by direct action at a distance. A central motivation was to make it possible to consider only actually existing charge values, in multiples of electron charge, rather than the arbitrarily small test charges necessary to define the electromagnetic field. The fact that elementary particles also have discrete mass values led Wheeler to the belief that, similarly, a continuous gravitational field is an unwarranted invention, which should only emerge in the limit of a large number of elementary particles. Another motivation for gravity as a direct inter-particle action was Mach's principle, that inertia is due to a direct interaction with the large masses in the universe. This idea was so persuasive that even after quantum electrodynamics made wheeler-Feynman action-at-a-distance unnecessary to get finite quantum results, Wheeler spent many years trying to eliminate the field aspect of gravity from Einstein's space-time theory and to do without space-time itself in favor of actually observable particles and light rays. The main problem was to find an action in terms of what he called the liaisons that light rays provided between world lines, which action would correspond to the General Relativity action in the space-time limit.

Even when such a liaison action could not be produced, the elements of liaison theory -- world lines and light rays -- made the gravitational fields of the General Relativity limit measurable and therefore physical.
 
Thus fields had returned, but did not immediately replace particles because of the intermediate possibility of particles as singularities in the field. For a brief period, regarding point particles as singularities in the field seemed attractive because their motion was seen (Einstein-Infeld-Hoffmann method) to be enforced by the field equations. But even more attractive than doing without separate equations of motion was avoiding singularities entirely and recognizing particles as special, nonsingular solutions of the field equations. With only qualitative ideas how such solutions could be constructed, Wheeler embarked in 1953 on a journey to Tokyo to give a lecture on the Problems of Elementary Particle Theory which is here given in its first English translation. To report on and distinguish both the established results and his own speculations he chose two spokesmen in the guise of well-known heroes of Japanese history. In their two views of elementary particle physics and its future the spokesman for "Fields First" promises exciting answers to questions like Mach's principle, the nature of mass, a theory without unexplained "natural" constants, an evaluation of the fine structure constant, and so forth. The elaboration of many of these ideas, such as mass from massless fields, is found two years later in the Geon paper, where Wheeler has totally returned to field theory and rejuvenates General Relativity with daring new applications.

\end{abstract}

In the mid-1950s, John Archibald Wheeler radically changed his research agenda, from a successful  and thoroughly mainstream career in nuclear physics to his trailblazing pursuit of general relativity, establishing in Princeton one of the most important hubs of the renaissance of relativity, and training a highly influential generation of American relativists.\footnote{Wheeler apparently resented the term ``relativist'', feeling that it implied too narrow a specialization on relativity \citep[p. 91]{bartusiak_2015_black}. We will still be using this term to denote a physicist with expertise in general relativity, making no value judgments about that physicist's expertise in other areas of physics.} Understanding Wheeler's turnaround will shed new light on the historical process that this volume focuses on. Yet, when speaking about his epochal shift, Wheeler tended to emphasize the metaphysical aspect, highlighting the transition from a particle to a field ontology and portraying it as an almost spiritual and definitely rather personal conversion: 

\begin{quote}
And of course nobody gets religion like a reformed drunkard. As I've often said about this subject, the fanaticism, if you would like to call it that, with which I pursued the opposite approach---that it's a pure field theory explanation of nature that one ought to work at---comes from having worked so hard at a pure particle explanation of what one sees.\footnote{Oral History Interview by Gloria Lubkin and Charles Weiner, conducted on 5 April 1967. \url{https://www.aip.org/history-programs/niels-bohr-library/oral-histories/4958}}
\end{quote}

In this paper, we wish to somewhat correct this highly attractive narrative of personal epiphany. We wish to show that it was the use of a specific methodology and the pursuit of a specific (albeit ambitious) research goal that ultimately led Wheeler into the (pardon the pun) field of general relativity.\footnote{Wheeler's transition has also been recently studied by Dean \citet{rickles_2018_geon}. That analysis focuses more on a before-after comparison, as well as the Geon paper and its aftermath, not on the details of Wheeler's transition, as we do in this paper. It thus makes good complementary reading.} Wheeler came to recognize general relativity as the central theoretical tool in pursuing a longstanding research program. We aim to show that Wheeler's turn to general relativity was thus a paradigmatic example of what Roberto Lalli, J\"urgen Renn, and one of the authors have called the recognition of the untapped potential of general relativity. Reconstructing Wheeler's intellectual biography in the years c.1935-1954 thus provides, besides its intrinsic interest, an explication of what this untapped potential, which was identified as the key epistemic factor in the renaissance of general relativity, precisely meant in the case of one of the central historical actors. Our reconstruction is based to a large part on the extensive archive of Wheeler's papers at the American Philosophical Society in Philadelphia, in particular on a set of notebooks that provide detailed insights into the development of Wheeler's thinking in the 1950s. 

The research goal that led Wheeler to general relativity was a radically reductionist one, of reducing the ever-growing number of elementary particles to one fundamental constituent, while still being able to explain and reproduce the multitude of observed particle properties and especially the still ill-understood nuclear interactions. Clearly, Wheeler was not alone in this research goal, which has been a central aspect of attempted ``final theories'' in physics, from Heisenberg's non-linear spinor theory\footnote{As analyzed in detail in the forthcoming volume ``Heisenberg's 1958 \emph{Weltformel} and the roots of post-empirical physics'' of the \emph{Springer Briefs in the History of Science and Technology} by one of the authors (AB). That book and this chapter represent key case studies for work of the research group \emph{Historical Epistemology of the Final Theory Program} at the Max Planck Institute for the History of Science.} to contemporary string theory. What made Wheeler's program so radically different, was that the proposed fundamental constituent was not to be a still unobserved microscopic element, but rather a known entity, one whose properties were established and codified in generally accepted physical theories. This methodology is what Wheeler would later call ``daring conservatism'':  taking well-established theories and (daringly) applying (and trusting) them far beyond their traditional domain, e.g., by applying them at microscopic length scales where they could not possibly have been experimentally confirmed. The theory that Wheeler ultimately selected as the key element in his reductionist program was, of course, general relativity.

In Section \ref{sec:feynman}, we will reconstruct the origins of Wheeler's program, which initially had nothing to do with gravity, instead relying solely on electrons and their electromagnetic interactions as a substitute for the speculative theories of nuclear interaction of the day. At the time, renormalization methods had not yet been developed and quantum electrodynamics (QED) was still a highly defective theory, beset by infinities. Wheeler had convinced himself that the  extrapolation of electrodynamics to the microscopic (and then the nuclear) realm would necessitate its reformulation in terms of an action-at-a-distance (AAD) theory (Wheeler-Feynman electrodynamics). Wheeler's hopes that electrodynamics could provide the key to the nuclear interactions began to falter in the late 1940s. He did not, however, abandon the idea of replacing field theories with interactions at a distance, which he now began to apply to gravitation (Section \ref{sec:AAD}). While Wheeler had originally hoped to construct an AAD theory of gravity from scratch, he became increasingly aware that he would have to take general relativity as a starting point. There followed an intense period of studying general relativity, also during the first course he taught on the subject in the academic year 1952/53, as we will discuss in Section \ref{sec:course}. 

His study of general relativity led him to appreciate the great potential of general relativity and to include it alongside electromagnetism in the established foundations of physics he would allow himself to draw upon. He initially focused on one feature of general relativity in particular, namely the way in which it related the source-free field equations to the motion of point-like particles in that field. He consequently adopted a new physical picture, where particles were described by singularities in the fields. Fields were thereby reinstated into his worldview, increasingly becoming the primary entities (Section \ref{sec:singularities}). As this research program matured, he began presenting it, along with his methodology of daring conservatism, in public lectures, first in a lecture in Japan in the fall of 1953. This Japanese connection had Wheeler contemplate the name ``Tokyo program'' for his research agenda, a name he soon dropped, but which we have resurrected for the title of this paper. In early 1954, he finally abandoned the last vestiges of his particle program as he discovered a new way of drawing on the untapped potential of general relativity: to construct particles as regular (non-singular) field configurations. This led to his famous geon paper \citep{wheeler_1955_geons} and the establishment of a general relativity as the central research focus of Wheeler and a new generation of his students. We discuss this final shift to a pure field theory and his explicit reflections on the daring conservatism program in Section \ref{sec:fields}, before we conclude in Section \ref{sec:conclusions}.

\section{The Great White Hope}
\label{sec:feynman}

The origins of John Wheeler's long and winding path to general relativity are hard to trace; all we have to go by are Wheeler's later recollections of the pipe dreams of a physicist in his mid-20s, or, as Wheeler liked to put it, his ``great white hopes.'' Of these there were two, which Wheeler recounted in two interviews, conducted by Charles Weiner and Gloria Lubkin (WL) in 1967 and by Finn Aaserud (FA) in 1988 (Session I),\footnote{\url{https://www.aip.org/history-programs/niels-bohr-library/oral-histories/5063-1}} respectively.

The first great white hope was electrons. After the discovery of the positron in 1932, the theory of electrons and positrons based on the Dirac equation, ``pair theory'', had been worked out by Dirac, Heisenberg, Oppenheimer and Furry. Wheeler felt that pair theory offered ``mechanisms for binding electrons in very small regions of space that never got a thorough discussion'' (WL), and that electrons might well be present in the nucleus after all (an assumption that had been dropped after the discovery of the neutron); in fact that electrons and positrons may form its fundamental constituents.

The second great white hope was scattering, which was to be viewed as the fundamental process from which all other characteristics of (primarily nuclear) interactions were to be derived. Both great white hopes were (at least in hindsight) also imbued with snappy and parallel slogans \citep{wheeler_1989_the-young}, ``Everything as Electrons'' and ``Everything as Scattering,'' and even if these precise titles are not actually contemporary, they do show an essential characteristic of Wheeler's thinking: An extreme reductionism, a reduction to simple, catchy thoughts and a very small number of fundamental building blocks, a radical Ockhamism if you will. In particular, we see here what Wheeler would later call ``daring conservatism'': Taking (an element of) a well-established theory, and trying to use it beyond its usual domain of applicability, i.e., electrons in the nucleus or scattering theory to describe stationary states. But this unique approach did not show at the time. Feynman would later remark (perhaps apocryphally) that: 

\begin{quote}
Some people think Wheeler's gotten crazy in his later years, but he's always been crazy.\footnote{Cited in \citep{overbye_2002_peering} and in \citep{wong_2008_remembering}. In the former, Kip Thorne is given as the reference for Feynman's supposed remark.}
\end{quote}

While the later Wheeler would happily have publicized catchy slogans for crazy ideas, such as ``everything as electrons,'' there are no outward indications of his grand vision at the time. As he would remark in the interview with Weiner and Lubkin:

\begin{quote}
Nobody was as crazy as I was, to think that you could explain everything in terms of electrons. And this I think illustrates a weakness of my approach at that time, to have this secret hope nursed internally and talk about it occasionally with close friends but not feeling particularly at ease about bringing it out on a public platform...
\end{quote}

Similarly, a talk \citep{wheeler_1934_interaction} he gave on (alpha particle) scattering at the APS meeting in Washington, DC, gave no indication (at least from the extant abstract) of the central role he was envisioning for scattering in fundamental physics, nor did his central paper on the subject, in which he famously introduced the S-Matrix \citep{wheeler_1937_on-the-mathematical}.

It would take the meeting with a fellow eccentric to tickle Wheeler into presenting hints of his crazy ideas on ``everything as electrons.''\footnote{We shall have no more to say about ``Everything as Scattering,'' which was here mentioned only to illustrate the early traces of daring conservatism in Wheeler's thinking. This notion was very influential in Feynman's later diagrammatic formulation of renormalized QED, which did effectively become a pure scattering theory, see \citep{blum_2017_the-state}.} We refer, of course, to his PhD student, Richard Feynman, with whom he worked out what would later be known as Wheeler-Feynman electrodynamics \citep{wheeler_1945_interaction}. In this theory, the field-mediated electromagnetic interaction of Maxwell's theory is replaced by a direct interaction at a (spatial) distance between charged point-like particles. This interaction is not instantaneous (in time), but is rather the sum of a retarded and an advanced component, corresponding to the two possible solutions of Maxwell's field equations and ensuring compliance with the special theory of relativity. When a given charged point particle (electron) exerts an advanced force on the other electrons in the universe, it will experience a retarded back-reaction, which will in fact be instantaneous. Imposing the ``absorber boundary condition'' (which in the corresponding field theory would imply that there are enough electrons in the universe to absorb all outgoing radiation, so that there is no radiation ``escaping to infinity'') then ensures that this instantaneous back-reaction is equal to the radiation reaction of the usual field theory. This then not only implies empirical equivalence with field theory, but also eliminates the possible difficulties with causality an advanced interaction might otherwise suggest.

Wheeler and Feynman give conflicting stories concerning the origins of their joint work on action-at-a-distance electrodynamics. Feynman in his 1965 Nobel lecture relates how, as an undergraduate, he had hit upon the idea of replacing the electromagnetic field with action-at-a-distance in order to eliminate the divergences of quantum electrodynamics (QED), how he had then learned as a graduate student that one cannot explain radiation reaction in this way, how he had tried to get radiation reaction into his action-at-a-distance framework as the backreaction of electrons, and how he had then presented this idea to Wheeler. Upon which Wheeler ``then went on to give a lecture as though he had worked this all out before and was completely prepared, but he had not, he worked it out as he went along,'' a lecture that ended in the conclusion that one would have to take into account advanced solutions of Maxwell's equations, in order to get an immediate radiation reaction. From this conversation then grew, as Feynman recalls it, their joint work on action-at-a-distance electrodynamics.

Wheeler's version of events is considerably younger, first published only a year after Feynman's death \citep{wheeler_1989_the-young}. He recalls how Feynman had expressed some interest in Wheeler's idea of everything as electrons, how Wheeler had ``animated by the concept of everything as electrons' worked out one Sunday (up to a factor of 2) how one might get radiation reaction as a backreaction from the absorber even in a theory without fields, and how he had then presented his caculation to Feynman, who was able to sort out the missing factor of 2. But priority issues are not our concern here. In fact, in this case not only is the issue undecidable, the two stories are not even entirely incompatible: If we assume that both Feynman and Wheeler omitted substantial parts of the story, the two accounts might actually be merged together to form a coherent narrative.

For our story, another aspect evidenced by the two different accounts is much more important: It clearly shows that Wheeler and Feynman came to action-at-a-distance electrodynamics with very different motivations. For Feynman it was the divergence difficulties of QED, an issue which had somewhat dropped out of fashion in the late 1930s due to the great interest in meson physics, but an issue that would soon resurface in the late 1940s and ultimately earn Feynman the Nobel Prize. For Wheeler, it was his great white hope, his ``everything as electrons.''

According to Feynman's recollections, it was he who got to write the first draft of their joint paper:

\begin{quote}
[Wheeler] asked me to write the paper - I wrote this thing up in 27 pages, which we could have sent to a journal, but he began to think, ``No, it's too great a business, we'll write it good.'' And that of course made delays, and got interrupted with the war, and he got it so big that it was five parts - the whole reorientation of physics from a different point of view. I never went along with him on that. I mean, you know, with the idea that it's so marvelous, it's a reorientation of physics, you have to write five papers, and all of physics is turned upside down. But I felt that 27 pages were what it deserved. This was written mostly by him. See, it was a rewrite of the 27 pages, so to speak. I wouldn't say a rewrite because he didn't use the 27 pages as a basis, but the same ideas are developed, which I tried to write much more briefly, and which he tried to write in an historical context, about the arguments of Tetrode and Einstein - you see, it's a relatively long thing, and I didn't really write it, you understand.
\end{quote}

The manuscript that Feynman is referring to is probably identical with an untitled manuscript from 1941 in the Feynman Papers at Caltech (Box 6, Folder 1).\footnote{The number of pages that this manuscript has depends on how exactly one counts, e.g., if one includes handwritten inserts, typed pages containing just one extraneous paragraph, and figure captions. A case can certainly be made for 27.} To assess the difference in tone, compare the first sentence of Feynman's draft:

\begin{quote}
The attempts to develop a satisfactory scheme of quantum electrodynamics have met with several difficulties, some of which are found not to be a result of the process of quantization, but to be contained in the classical electron theory itself.
\end{quote}

with the opening of the final, printed version:

\begin{quote}
It was the 19th of March in 1845 when Gauss described the conception of an action at a distance propagated with a finite velocity, the natural generalization to electrodynamics of the view of force so fruitfully applied by Newton and his followers. In the century between then and now what obstacle has discouraged the general use of this conception in the study of nature?
\end{quote}

Wheeler's grand vision clearly came out in the historical pathos. But as far as the motivation provided for reformulating classical electron theory in terms of action-at-a-distance, Wheeler fully adopted Feynman's motivation of solving the difficulties of QED. There is no hint of Wheeler's great white hope and so we need to ask the question, what did Wheeler mean when decades later he referred to his work as being ``animated by the concept of `everything as electrons'? Now, one might simply take this to mean that he was looking for a purely particulate description of nature, only electrons, no fields, and that action-at-a-distance was the implementation of this program. There is certainly something to this, and in his later recollections Wheeler certainly emphasized the ``everything as particles'' aspect of his early period, contrasting it with his later pure field approach. 

But there is another aspect that appears to have been equally important, if not more so. Wheeler's program was specifically focused on electrons, and especially on their continuing role in the nucleus. It was thus not simply concerned with the electrons as generic particles, but with electrons as a very specific kind of particles, distinct from nucleons or mesons. Now, he was well aware that there were significant indications that electromagnetically interacting electrons were not the primary constituents of the nucleus, or even present there at all: The short range of the nuclear interactions and the huge kinetic energies obtained by electrons thus confined were the indications most typically cited. It seems, now, that Wheeler hoped that the action-at-a-distance formulation might help to put aside these objections to ``everything as electrons,'' as he also later stated in his autobiography \citep[p. 164-165]{wheeler_1998_geons}:

\begin{quote}
I had another motivation as well for pursuing action at a distance, for I clung to my hope that all of the matter in the world could be reduced to electrons and positrons. Yet I knew that if an electron and a positron were to be crowded together in subnuclear dimensions, some way would have to be found to get around the prediction of conventional theory that they would quickly radiate away their energy in the form of electromagnetic fields. Perhaps, I thought, an action-at-a-distance version of electromagnetic theory - one without fields - might explain the suppression of such radiation and permit the particles to live happily in such a confined space.
\end{quote}

There are only vague hints at this in the published paper with Feynman, which contains a section on ``Advanced effects associated with incomplete absorption,'' which discusses the physics of electrons in an incompletely absorbing cavity. This may be interpreted as Wheeler thinking about electrons in the nucleus, but this possibility is not mentioned explicitly. Stronger hints can be found in a lecture he gave, several months after the publication of the first Wheeler-Feynman paper, at a symposium of the American Philosophical Society on ``Atomic Energy and Its Implications.`` This symposium, conducted in November 1945, only three months after the use of atomic bombs on Hiroshima and Nagasaki, was a rather serious affair ``devoted to the atomic bomb'' (as per the opening words of Henry DeWolf Smyth), featuring talks by J. Robert Oppenheimer on ``Atomic Weapons'', Joseph H. Willits on ``Social Adjustments to Atomic Energy'' and Irving Langmuir on ``World Control of Atomic Energy'' (Proceedings of the American Philosophical Society, Vol. 90, No. 1, January 1946). Wheeler recalled that he ``didn't really want to talk about atomic energy'' but rather ``about what lay beyond it.'' \citep[p. 168]{wheeler_1998_geons} So, inspired by his collaboration with Feynman and eager to do fundamental physics after the long war work, he chose this unusual venue to provide first hints on his grander schemes in a talk entitled ``Problems and Prospects in Elementary Particle Research''.

The hints towards his hope that action-at-a-distance electrodynamics might revolutionize nuclear physics are still very vague, merely by association \citep[p. 45-46]{wheeler_1946_problems}:

\begin{quote}
[T]he theory of [nuclear] mesons, may be said to be at present in a state of free experimentation with ideas and great uncertainty as to principle, both because of the incompleteness of our present experimental picture and because of difficulty in tying the proposed hypotheses to already existing theories. [...] The difficulties of inventing a new theory on the basis of  incomplete experimental evidence suggest that one possibility acceptable at this time is the conservative one of extending the range of applicability of already existing and well-established theories.\\
The second theory whose problems we consider is therefore the formalism of electron-positron pairs.
\end{quote}

We see here the notion of daring conservatism (without that name yet) appear in print for the first time, but as to how it is to be applied, there is nothing to go by but a lone and easily overlooked ``therefore.'' Wheeler would focus on cosmic ray physics in the following years, setting aside once again his great white hope (with the exception of some work with his student Gilbert Plass on counting degrees of freedom and thus doing thermodynamics - Planck's law - in an action-at-a-distance theory). But in June 1946, he submitted an application to the Guggenheim Foundation to pursue a project on an ``Analysis of the Problem of Measurement in Electron Theory.'' Through this scholarship, he hoped to obtain a leave of absence from Princeton to work on his foundational ideas. And he could think of no one better suited to help him in this endeavor than his old mentor Niels Bohr. As he wrote in his application:

\begin{quote}
Only a start could be made in the [...] program by the applicant himself and in collaboration with his student, R.P. Feynman; the war stopped further work. [...]
The primary reason for this proposal is the feeling that the stage has now come in the theoretical work where new concepts and points of view are essential. To develop and test such points of view it appears that by far the most effective course is to take up a close association with Niels Bohr. The writer's association with this scientist has convinced him that if there is hope of making advancement in the fundamental problems outlined above, this hope is best justified by Bohr's ability to see into the future, his courage and judgment in considering and testing new concepts. [...] [T]he Applicant therefore is proposing to go to work with Professor Bohr in Denmark.
\end{quote}

Even in this proposal, Wheeler was still rather guarded about what he wanted to do exactly. The proposed title recalled Bohr and Rosenfeld's paper on the Problem of Measurement in QED \citep{bohr_1933_zur-frage}, which to Wheeler was the model for an incisive theoretical analysis of current theory \citep{hartz_2015_uses}, and which he referred to as ``a classic paper'' in his application. And when listing the tasks for his project, Wheeler remained excessively general:

\begin{quote}
(1) The consistent formulation of the mathematical formalism of the theory of electrons and positrons.\\
(2) A definition in terms of idealized experiments of the possibilities of measurement in the theory of the electrons and positrons.\\
(3) The interpretation of these idealized experiments in terms of the formalism - and of the formalism in terms of the idealized experiments.
\end{quote}

Consequently, Eugene Wigner, who acted as one of the references for Wheeler's application, was also rather non-committal about his views on Wheeler's project (received 24 June 1946):

\begin{quote}
As to his proposed study, I must admit that I find it quite impossible to make any predictions. [...] I know from personal contacts that Professor Wheeler is most deeply interested in the project which he outlined.
\end{quote}

In any case, while Wheeler's application was granted only five days after it was received (and five days before even receiving Wigner's letter of reference), Wheeler could not use the grant as soon as he had planned. As he wrote to Henry Allan Moe of the Guggenheim Foundation on 2 July 1946:

\begin{quote}
I have now learned on my return to Princeton that there is a distinct possibility that the six-month leave of absence granted by the University to Professor Wigner may have to be extended through the second term of 1946-47 to allow him to accomplish most effectively his task to give a new direction to the work of the Oak Ridge Laboratory. I am afraid after examination of the situation with the Department here that it would create embarrassing difficulties for the work now underway to have me gone for the whole of the Spring term. However, it appears that I may be able to count on a period May 15 - October 1, 1947. Even this range of time can, however, not now be made entirely definite. [...] I am sorry to have to report to you that the situation is in this uncertain state. I should like to have your advice as to how I can best take into account these difficulties in a manner acceptable to the Foundation.
\end{quote}

There was no problem. Moe replied on 5 July 1946:

\begin{quote}
It is a pleasure to welcome you to the company of Guggenheim Fellows -- the distinguished company as I think.\\
As to the date of starting your Fellowship, there's no need to settle that now. [...] You may count on us to ``play ball.''
\end{quote}

After some back and forth, Wheeler finally received his scholarship for the period from 1 July 1949 to 30 June 1950 (i.e., also for an extended time period). For family reasons \citep[p. 183]{wheeler_1998_geons}, he also opted to make Paris his home during his stay in Europe, rather than Copenhagen. And even though he did visit Copenhagen on several occasions during his time in Europe, much of his work with Bohr was focused not on electron-positron theory, but rather on completing a paper with David Hill on the collective model of the nucleus, which Bohr ended up not co-signing (see Acknowledgments in \citep{hill_1953_nuclear}). 

One reason for this appears to have been a major advance that occurred in the years from 1946 to 1949: The problem of the electromagnetic self-energy of the electron had been solved, among others by Feynman, without the need to eliminate the self-interaction of the electron. Wheeler had always presented the elimination of the self-interaction as the prime motivation for action-at-a-distance. Consequently, when writing an updated version of his 1945 talk in 1949, he duly acknowledged the recent advances in quantum electrodynamics and used a much more cautious language when talking about Wheeler-Feynman electrodynamics. In 1945, he had concluded his brief elaborations on Wheeler-Feynman electrodynamics with the words:

\begin{quote}
It is too soon to say whether the translation of the revised classical theory into quantum mechanics will remove the outstanding divergences. To test this point is an important problem for the future.
\end{quote}

In 1949 he wrote:

\begin{quote}
Naturally the ultimate complete equivalence of this approach to the usual field theoretical treatment makes it clear that nothing new can result so long as this equivalence is strictly maintained. What may come out by changes and reinterpretations of the existing theory of action at a distance is uncertain \citep{wheeler_1949_elementary}.
\end{quote}

That he did not yet give up action at a distance entirely at this time (as Feynman did, see \citep{blum_2017_the-state}) is clearly due to his great white hope of constructing the atomic nucleus from electrons, a conjecture that was largely unaffected by the advances in QED. Indeed, we have one bit of evidence that Wheeler did in fact discuss this idea with Bohr during his stay in Europe, a letter sent to Bohr from Paris on 21 January 1950,\footnote{Archives for the History of Quantum Physics, Bohr Scientific Correspondence, Microfilm 34.} announcing his arrival in Copenhagen for the 27th. This letter is primarily concerned with their joint work on the collective nucleus, but then, on page two, Wheeler brings in ``everything as electrons'':

\begin{quote}
The other problem, about which I am very anxious to get your opinion, is the question: is it possible to exclude a picture of elementary particle constitution based entirely on positive and negative electrons?
\end{quote}

Wheeler discussed his hope how half-retarded, half-advanced action-at-a-distance might solve the problems usually brought forth:

\begin{quote}
(1) \emph{Localizability}. To localize an electron in a distance of $\approx e^2/mc^2$ it is sufficient to have a potential well of radius $(e^2/mc^2)$ and depth $(137)^2 mc^2$ [...]\\
(4) \emph{Source of potential}. Acceleration and velocity of an electron bound in such a potential are so great that electrostatic forces are negligible in comparison with radiative forces. Radiational transfer of energy to the outer world is itself negligible owing to the symmetry of the charge-current distribution. The radiational forces within the system can therefore be considered in a good approximation as half-advanced, half-retarded [...]. [I]t does not seem impossible to suppose that the electronic system generates a self-consistent potential, in which however the correlations between movements of interacting electrons must have an altogether dominating importance, in contrast to the atomic case.
\end{quote}

All this was very qualitative. And even if Wheeler's hopes concerning the nuclear binding potential could be realized, he was aware of the fact that there was another difficulty, which he had only very vague ideas on how to address: That of spin and statistics. If a neutron, e.g., was really to be considered as consisting of electrons and positrons only, it would have to contain an equal number of electrons and positrons to ensure its neutrality, and thus would have to be an integer-spin boson. Wheeler was clearly interested in Bohr's judgment as to whether the project was worth pursuing despite this apparently insurmountable difficulty:

\begin{quote}
Though I joke with you about my heresies, I am trying to be just as honest and as open as I can about the elementary particle problem. I know that there is no one who has your insight. So it will be a great privilege to talk with you over these and other problems...
\end{quote}

As this letter was discussed during Wheeler's visit to Copenhagen, we have no evidence as to Bohr's take on the matter. Soon after his visit to Copenhagen, Wheeler cut his sabbatical short to work on the hydrogen bomb. When Wheeler returned to foundational research after the hydrogen bomb interlude, his focus was now on a different, but related problem: action-at-a-distance gravity.

\section{Action-at-a-distance Gravity}
\label{sec:AAD}

The idea of action-at-a-distance gravity can be traced back to 1941, when Wheeler and Feynman had worked out the basics of their absorber theory and Feynman first presented their joint work at a seminar in Princeton. Feynman later recalled that Einstein had remarked \citep[p. 80]{feynman_1985_surely}:

\begin{quote}
I find [...] that it would be very difficult to make a corresponding theory for gravitational interaction...
\end{quote}

Wheeler appears to have been immediately intrigued by this challenge. Although soon mainly occupied by war work, he wrote a letter to Einstein on 3 November 1943 (Einstein Papers), requesting a meeting with Einstein to discuss ``where the force of gravitation fits into the point of view'' of action-at-a-distance theory. The meeting took place on 14 November in Wilmington, Delaware  (Letter from Einstein to Wheeler, 6 November 1943, Einstein Papers) and we can surmise some of its content from a letter which Wheeler wrote to Einstein after the meeting, on 2 December. Wheeler had developed a general framework for action-at-a-distance theories, which he referred to as ``the theory of world lines.'' In this framework not only was there no more talk of fields, there was even ``no reference to the concept of a space-time continuum.'' All that was left were the world lines of individual particles $a$, $b$, etc., the points on which were identified by  some parameter, $\alpha$, $\beta$, etc. The exact parameterization of the world lines was to be considered arbitrary, and all statements on physics were supposed to be independent of it.

The dynamics were now determined by functions that connect points on two worldlines. In his letter to Einstein, Wheeler referred to these functions as light cones, which most clearly reveals their physical interpretation. In his later notes, he would mainly use the term liaison. And when he finally wrote a paper on AAD gravity some 50 years later, together with Daniel Wesley, he called them associators \citep{wesley_2003_towards-an-action-at-a-distance}.\footnote{One reason for Wheeler's return to action-at-a-distance gravity after so many years appears to have been that it had by then become abundantly clear that quantum gravity would suffer from similar divergence difficulties as QED, difficulties that in electrodynamics the Wheeler-Feynman absorber theory had aimed to solve. Since these difficulties could not be solved by renormalization in the case of gravity, it appeared attractive to revisit the old AAD formulation. This was pointed out to us Daniel Wesley, who co-authored the AAD gravity paper with Wheeler as an undergraduate student, in an email to one of the authors (AB), 11 April 2019.} We will be referring to them as liaisons throughout this paper, as this is the term that Wheeler used for most of the period under study.\footnote{We will also be using a notation that Wheeler introduced only after his letter to Einstein. This is additionally motivated by the fact that reproducing Wheeler's notation in the Einstein letter presents somewhat of a typesetting challenge.}

The liaison function $\alpha_+ (\beta) $ then returns the point $\alpha$ that is on the forward lightcone of $\beta$.\footnote{\label{fn:conjugate} In general, the forward lightcone might intersect the other world line more than once. The liaison was supposed to be a single-valued function that singles out one of those points and thus constituted an object somewhat more restricted than a lightcone.} Similarly, $\beta_+ (\alpha) $ returns the point $\beta$ that is on the forward light cone of $\alpha$. Wheeler also introduced liaisons $\alpha_-$, $\beta_-$ for the backward lightcone, which are the inverses of $\beta_+$ and $\alpha_+$ respectively. Wheeler now proposed to Einstein that one could construct physics from these functions by setting up non-trivial, parameterization-independent relations. He pitched the expression
 
 \begin{equation}
 \alpha' = \alpha_- (\gamma_+ (\beta_+ (\alpha))),
 \end{equation}
 
where $\alpha'$ is the  point on world line $a$ that one reaches by moving along the forward light cone from a point $\alpha$ to world line $b$, then further on the forward light cone to world line $c$ and then on the backward light cone back to world line $a$. As can easily be seen by imagining all three particles as being at rest in the (not yet constructed) three-dimensional background space, if $\alpha'=\alpha$ the three particles are on one line in three-space. If $\alpha' \neq \alpha$ they are not. One could thus, merely from three one-dimensional world lines and the liaisons between them, distinguish between a line and a triangle through the (indeed parameterization-independent) statement that $\alpha$ and $\alpha'$ are equal or unequal, respectively. Wheeler further conjectured:
 
 \begin{quote}
 With a number of particles greater than three, one can build up more complex geometrical concepts.
 \end{quote}
 
Apparently, Wheeler had hoped that in this manner one would be able to construct a theory that would be fully equivalent to general relativity, just as Wheeler-Feynman absorber theory was equivalent to Maxwell. In their discussion, Wheeler and Einstein appear to have focused on a surprisingly specific difficulty with such an equivalence: A universe with two particles. In the world line theory, Wheeler asserted, ``no physics at all is possible'' in such a setup, because the only expression one can study, namely $\alpha_- (\beta_+(\alpha))$, is trivially the identity, since $\alpha_-$ is \emph{defined} to be the inverse of $\beta_+$.\footnote{In general, it is of course possible to first move along the forward lightcone from worldline $a$ to worldline $b$ and then back, along the backward lightcone, to worldline $a$ and arrive at a point different from the point one started out from, when conjugate points are involved. In the liaison framework this issue is avoided: As mentioned in footnote \ref{fn:conjugate}, the liaison has to be a single-valued function and thus it was natural on many levels for Wheeler to simply define the backward liaisons as the inverses of the forward ones.} In general relativity, on the other hand, there were, as Einstein pointed out, two-body solutions with a rich, non-trivial four-dimensional geometry.\footnote{The two-body solutions by Weyl and Levi-Civit\`a that Einstein was referring to are discussed in detail in an editorial footnote of the Collected Papers of Albert Einstein \citep[p. 437]{buchwald_2018_the-collected}.} There was thus, as Wheeler remarked, ``an apparent discrepancy between the general theory of relativity and the general theory of action at a distance.''

In his letter to Einstein, Wheeler began to develop an understanding, which he would further develop over the course of the following years, that his world line theory was in fact right in implying that there should be ``no physics'' in a two-body universe. Wheeler would present first hints of his search for an AAD formulation of gravity in several talks given in the first postwar years, such as in his programmatic 1945 talk at the American Philosophical Society, already mentioned earlier, where he remarked:

\begin{quote}
Just as the proper recognition of [...] atomicity requires in the electromagnetic theory a modification in the use of the field concept equivalent to the introduction of the concept of action at a distance, so it would appear that in the gravitational theory we should be able in principle to dispense with the concepts of space and time and take as the basis of our description of nature the elementary concepts of world line and light cone.
\end{quote}

But his stance on the two-body problem was only spelled out in the revised 1949 version of the talk. In defense of world line theory, Wheeler had adopted a relationalist view of space and time, in the tradition of Ernst Mach who had formulated such ideas in the late 19th century in opposition to the Newtonian notions of absolute space and time. Mach's ideas, in the shape of the more or less formalized ``Mach's Principle'' of the relativity of inertia, had played an important role in Einstein's application of general relativity to cosmology \citep{smeenk_2014_einsteins}. But the existence of two-body solutions, Wheeler now argued, called into question the validity of Mach's principle in general relativity. In a strictly relationalist theory, an idealized two-body problem would be non-dynamical: There was only one length scale to relate things to, the distance between the two point masses, and hence that distance should not be representable as changing, since there was nothing that its change could be measured against. Any statement that general relativity claimed to be able to make about the time evolution of the distance between the two bodies was thus empty.

While the centrality of the two-body problem was not to persist (already in the 1949 lecture, Wheeler voiced his doubts whether the paradox would necessarily be resolvable by action-at-a-distance gravity), the years 1943-49 see gravitation moving to the center of Wheeler's attention, with new interesting paths of inquiry popping up, such as the unresolved problem of gravitational radiation reaction. While his talks of the period, as mentioned earlier, contained somewhat defeatist language concerning the prospect of Wheeler-Feynman electrodynamics, this is not the case for action-at-a-distance gravity. As his hopes for explaining the nuclear forces through electrodynamics were waning, gravity was increasingly presenting itself as a worthwhile field of study. Wheeler was beginning to realize the untapped potential of general relativity, though it must be admitted that he does not yet appear to have had a definite program for using this potential. The most specific part of his new interest in gravitational theory was the focus on the role of Mach's principle (beyond the particular case of the two-body universe) in establishing the relation between the world line theory and general relativity. This became, as we shall see, a central theme in Wheeler's thinking about gravitation.

Wheeler appears to have worked on AAD gravity quite a bit during his 1949/50 stay in Europe. In a letter to Gregory Breit of 28 December 1949 (Gregory Breit Papers, Yale University Archives), he wrote:

\begin{quote}
I am working quietly, sometimes on the reconciliation of the individual particle model of the nucleus and the liquid drop model [i.e., collective models of the nucleus], sometimes constructing a description of nature which makes no use of the concepts of space and time (analogue in gravitation theory of electromagnetic action-at-a-distance).
\end{quote}

Similarly in a letter to Feynman of 10 November 1949 (Feynman Papers, Caltech). And also in the letter to Bohr of 21 January 1950, already cited earlier, Wheeler talked about action-at-a-distance gravity, explicitly connecting it with Machian ideas of making ``force dependent upon the number of particles in the universe'' and mentioning another letter (not extant) to Wilhelm Magnus, mathematician at the Courant Institute, who had provided Wheeler ``with some information about one of the group theoretical aspects of the problem.'' In later recollections, Wheeler even misremembered that he had proposed ``doing similar ideas [to Wheeler-Feynman] for gravitation theory'' (FA) already in his application to the Guggenheim Foundation in 1946. But at the time, his main focus was still on electromagnetism. With the success of renormalized QED, electrodynamics was on its way out for Wheeler, and Bohr appears to have disabused him of the last vestiges of ``everything as electrons.'' But gravitation, in Wheeler's mind, was up and coming! In April 1951, he returned to Princeton from Los Alamos, and while still chiefly concerned with the work on the hydrogen bomb \citep[p. 218]{wheeler_1998_geons}, he did find some time to ponder these foundational questions. In a notebook entitled ``Action at a distance I'', we find an entry dated 10 November 1951, in which Wheeler considers two different pathways to AAD gravity:

\begin{quote}
Can thus work towards desired [action] principle from either one of two directions---(1) math. convenience + naturalness; (2) correspondence.
\end{quote}

And at this point the second pathway, establishing the theory through correspondence with the field theory of general relativity, clearly seemed the less favored, especially since the correspondence could not be exact for small numbers of particles, as he had established in his discussions with Einstein several years earlier. A few lines above the remark just quoted, Wheeler had written

\begin{quote}
But should satisfy the principle of correspondence to ordinary general relativity in the limit of infinitely many infinitely small masses (continuous mass distribution).
\end{quote}

only to then qualify this remark by a ``probably'' inserted after ``should.'' So how did Wheeler's pursuit of AAD gravity along the lines of mathematical convenience and naturalness look? Still using the liaisons as his central dynamical variables, Wheeler's idea was now to set up an action principle (similar to the Fokker action in Wheeler-Feynman electrodynamics), written as an integral over world line parameters, the integrand being some function of the liaisons. When varied, this action would return differential equations for determining the liaison functions. As to how this action should look, Wheeler thought that it might involve counting closed cycles of liaisons, as this could provide a notion of local world line density without having to invoke an underlying space-time. 

While these general ideas appear to have been present already in late 1951, Wheeler did not elaborate on it any further in his AAD notebook for quite a while. The next entry dealing with liaison theory dates from 17 March 1953. It would appear that Wheeler's purely mathematical approach turned out to be inadequate. For in 1952, he switched gears and started to engage general relativity head-on.

\section{Teaching Relativity}
\label{sec:course}

Embarking on his study of general relativity and the corresponding action-at-a-distance formulation, Wheeler asked to teach a course on relativity at Princeton in the academic year 1952/53. His request was granted on 6 May 1952, the day on which Wheeler began his first in a long series of notebooks on Relativity.\footnote{We will be citing frequently from Wheeler's first two relativity notebooks, which we will be abbreviating as WR1 and WR2, respectively. These notebooks are to be found in the John Wheeler Papers, held at the American Philosophical Society in Philadelphia, Section V, Volumes 39 and 40. } It is somewhat surprising that Wheeler was the first to teach a dedicated relativity course at the Princeton physics department. After all, Wheeler himself had clearly profited from the fact that Princeton was  \emph{the} center for relativity in the US at the time. We have already mentioned his personal discussions with Einstein. And in his first writing on relativity, the 1945 talk at the American Philosophical Society, every author cited in the section on gravitation, aside from Ernst Mach, worked in Princeton. However, they all worked at the Institute of Advanced Study, resulting in a great divide between the accumulated expertise on relativity at Princeton and the lack of relativity teaching. There was expertise at the university and Wheeler tapped into that as well, as recalled by Churchill Eisenhart, son of Princeton professor Luther Eisenhart:

\begin{quote}
As I understand it, after [Luther Eisenhart] retired he and John Wheeler were working together at writing a book called Mathematics Essential for the Theory of Relativity. [...] Dad and Wheeler, as I understand it, were bringing together in their book the mathematics, from here and there in the various branches of mathematics, you need for the general field theory.\footnote{Interview on 10 July 1984 with Churchill Eisenhart conducted by William Aspray, available at \url{https://www.princeton.edu/mudd/finding_aids/mathoral/pmc09.htm}. In this interview, Churchill Eisenhart also recalls that the manuscript for Wheeler and Eisenhart's book disappeared under mysterious circumstances after Luther Eisenhart's death.}
\end{quote}

Eisenhart retired in 1945, around the time that Wheeler began thinking about AAD gravitation.\footnote{For biographical information on Eisenhart, see \citep{lefschetz_1969_luther}.} But Eisenhart (like Valentine Bargmann, another Princeton University expert on differential geomertry) was a mathematician, and indeed up until Wheeler's initiative general relativity was only taught in the mathematics department at Princeton \citep{kaiser_1998_a-psi}.\footnote{Kaiser erroneously gives the year of Wheeler's first course as 1954/55. The 52/53 course, which is well documented by Wheeler's notebook, indeed did not yet show up in the Princeton course catalogue. A course on relativity by Wheeler is listed for 1953/54. This course catalogue had not been available to Kaiser at the time he wrote his paper.} So, despite the immense tradition and expertise that Wheeler could draw on in his exploration of general relativity, he was indeed the first one to teach it to Princeton physics graduate students.

Wheeler's notebook opens with his thoughts on his upcoming course:

\begin{quote}
5:55 pm. Learned from [Allen] Shenstone [then head of the Princeton Physics Department] 1/2 hour ago the great news that I can teach relativity next year. I wish to give the best possible course. To make the most of the opportunity, would be good to plan for a book on the subject. Points to be considered:\\
(1) a short introductory outline of the whole\\
(2) Emphasis on the Mach point of view \\
(3) Many tie-ups with other fields of physics. Mention these in class; in the book put them in the ends of chapters as examples
\end{quote}

The last two remarks are especially noteworthy. In remark (3), we can already see a recurring theme in Wheeler's later work on and in relativity, both intellectually and institutionally,\footnote{On Wheeler's role in ensuring that the institutionalization of research in general relativity would take place in the disciplinary context of physics, see \citep{lalli_2017_building}.} namely to establish general relativity as a physical theory, rather than a mathematical or philosophico-cosmological one. Here we see the decidedly pedagogical aspect of this theme, as only in this manner could it legitimately be taught to and applied by physics students. This emphasis should of course also be viewed in light of the predominantly mathematical tradition in relativity at Princeton University.

Wheeler's personal intellectual perspective on general relativity shows in remark (2), which hints at how strongly Wheeler's interest in General Relativity was at this point tied up with the prospect of an action-at-a-distance formulation, in which space-time disappears as an independent entity. Indeed, action-at-a-distance was a defining element in Wheeler's subsequent course as documented by his notes. The first term, which dealt primarily with special relativity, saw frequent references to the Wheeler-Feynman papers, including a long discussion of Wheeler-Feynman electrodynamics itself, stretching from December 1952 to January 1953.

Wheeler's general relativity class began in February 1953 with a first class meeting on 5 February in which topics for seminar reports were discussed. Here Wheeler had already honed in on some key physical problems, problems that would be defining elements of the upcoming renaissance of relativity: Gravitationally collapsing stars, gravitational radiation, and empirical cosmology. These three problems were joined, in a list of topics for seminar reports discussed in the first class meeting (WR1, p. 47), by an `Assessment of Unified Theories.'' Here, Wheeler was clearly attempting to make contact with the general relativity scene as it presented itself to him at Princeton, as witnessed by the list of references for this report, which included not only the obvious Einstein (specifically his latest paper, which had just appeared in the January issue of the physical review \citep{einstein_1953_a-comment}) but also work by Eisenhart, who had now, in retirement, turned to the study of non-symmetric metrics as they appeared in Einstein's Unified Field Theory \citep{eisenhart_1951_generalized}. The suggested topics for seminar reports are followed by an unsorted list of further topics Wheeler wanted to cover, which included both the ``Mach point of view'' and, immediately afterward, ``our particulate point of view.'' The list also contains the entry ``Variational principle and connection with quantum theory,'' a clear reminiscence to the least-action (in modern parlance: path integral) formulation of quantum mechanics that Feynman had developed precisely in the attempt to quantize Wheeler-Feynman electrodynamics. So, also in his list of topics for the second half of the course, we see the two central foci of physical problems (where Wheeler's identification of the central ones was clearly very influential) and of Wheeler-Feynman-Mach action-at-a-distance gravity, now joined by a rising interest in the idea of a unified field theory stimulated by the Princeton milieu.

Of course these questions were interrelated. The question of gravitational radiation, for example, was connected with the construction of an action-at-a-distance theory. The empirical adequacy of Wheeler-Feynman electrodynamics required imposing the so-called absorber condition that any electromagnetic radiation ultimately be absorbed, with nothing ever escaping to infinity. Is the gravitational world of GR ``non-absorptive'', Wheeler asks on the following page (WR1, p. 49), labeling it a ``\emph{very} vital question to look at.'' Wheeler was thus following an intellectual trajectory typical for the renaissance of GR: In pursuing a speculative extension of GR (action-at-a-distance in this case), he was forced to reflect on fundamental questions of GR proper (gravitational radiation). The question of absorber boundary conditions was really more of a side issue in this study of GR for ulterior purposes, however. As we saw in the last section, the central challenge for Wheeler was to construct a least action formalism for AAD gravity using liaisons. Wheeler pursued this program in parallel to teaching the course, and his lecture notes are consequently interspersed with research notes, initially focussing on the construction of liaison theory. In the following, we will focus almost exclusively on the research notes, leaving the exact reconstruction of the curriculum of Wheeler's course aside.

After his purely mathematical approach to this problem appears to have led nowhere, Wheeler's aim was now to construct liaison theory by studying its correspondence to regular GR. The first challenge here was to establish the locus of the correspondence, i.e., to identify the correct field quantity in GR that was to be reconstructed from the liaison formulation, the actual  ``gravitational field.'' Some two weeks into the course, Wheeler began to focus his attention on the Riemann tensor (WR1, p. 57). In an AAD theory, this would ultimately (via the liaisons) have to be reconstructed solely from the matter content (possibly merely in the form of singular worldlines), along with some sort of boundary conditions. Wheeler was thus led back to Einstein's original question concerning the realization of Mach's principle in GR: Is this sufficient to uniquely determine the Riemann tensor (up to coordinate transformations)? Wheeler ultimately reached a conclusion similar to that of \citet{einstein_1917_kosmologische}, namely ``that there is a one to one correspondence between mass distribution and metric only when space closes up on itself.'' (WR1, p. 103)

It should be noted that Wheeler was aware that this statement was merely a plausible conjecture: ``Any proof of uniqueness of case where metric is made to close up on itself? Very important question of principle.'' (WR1, p. 105) His main source for this conjecture was  \emph{The Meaning of Relativity} \citep{einstein_1953_the-meaning}. He continued to discuss this matter at Princeton with Weyl (WR1, p. 111), Wigner and von Neumann (WR1, p. 120), all of whom disagreed with Wheeler's assessment. Wheeler took this aversion to Mach's principle to also be a result of unfortunate formulation of the principle and gave his class the task of coming up with a better ``presentation of Mach's principle in 2 pages for an elementary physics student.'' (WR1, p. 135). Despite these difficulties, Mach's principle remained central to Wheeler's research program as it provided an analog of the Wheeler-Feynman absorber boundary conditions in general relativity. For some time, closure of the universe became an unquestionable fact to Wheeler, as he explained to his students:

\begin{quote}
Question raised in class whether mass density enough to permit open or closed universe, in view of expansion rate. Answer: [...] closure comes first, density knowledge too poor to permit proof of contradiction; closure so fundamental to whole Mach idea that in present state of knowledge think of density value having to yield precedence to Mach principle. (WR1, p. 104)
\end{quote}  

With the Riemann tensor identified as ``the field'' (WR1, p. 96), the possibility was now established to construct the action for liaison theory through correspondence with the usual field (Hilbert) action:
 
 \begin{quote}
 Set up an experimental procedure to get $R_{ijkl}$ locally by liaisons between a number of particles. In this way tie up $R_{ijkl}$ with liaison picture. Hence express $R$ in terms of local liaisons. Hence get variation principle in terms of local liaisons. (WR1, p. 89)
 \end{quote}
 
The ``experimental procedure'' was supposed to involve some sort of ``batting back and forth'' of light signals (WR1, p. 90) which would be a physical realization of the connection between two points established by a liaison. But Wheeler's attempts to tie up the Riemann tensor with the liaison picture ended inconclusively: He attempted to find the liaison function between two world lines from general relativity, where the light signals $\kappa$ travel from one particle world line to another on light-like geodesics, soon focusing on the limiting flat space case, where the two world lines $x$ and $\overline{x}$ are straight (WR1, p. 97), i.e. the system of equations:

\begin{eqnarray*}
x^i (s) & = & x^i (0) + s \left(\frac{dx^i}{ds} \right)_{s=0} \\
\overline{x}^i (\overline{s}) & = & \overline{x}^i (0) + \overline{s} \left(\frac{d\overline{x}^i}{d\overline{s}} \right)_{\overline{s}=0} \\
\kappa^i &=& \overline{x}^i - x^i \\
\kappa^{\alpha} \kappa_{\alpha} &=& 0
\end{eqnarray*}

which was supposed to give a relation between the parameters $s$ and $\overline{s}$, i.e., the liaison function giving for any point on one world line the point on the other one that lies on the first point's light cone. But even this simplified, non-gravitational trial calculation (27 March 1953; WR1, p. 99) ended inconclusively. His simple idea of obtaining the liaison action merely by translating the Hilbert action into liaison language faltered. Although now fully immersed in general relativity and tensor calculus, he returned to his original mathematical approach and began to pursue (8 April) a new approach to the liaison action, no longer based on counting cycles, but rather on counting the number of (forward) liaisons entering and exiting a given volume element, a setup inspired (as he himself remarked) by the neutron balance in a nuclear chain reaction (WR1, p. 114).

In all this searching, Wheeler was well aware that he was pursuing an entirely new style of doing physics. On 18 March 1953, in the margins of notes on liaison theory (AAD notebook), he remarked:

\begin{quote}
This mushy thinking may in end be much better, if less attractive, to present than the usual 1,2, 3 type of argument with which one at the end so often presents his special conclusions.
\end{quote}

What was driving him down this road of ``mushy thinking'' appears to have been the feeling of pursuing something grand, the ``great white hope'' feeling for which we here have the first contemporary archival evidence. Framing to himself his attempt to eliminate space and time, he wrote 

\begin{quote} 
Undoing work of early man, that theoretical physicist who left no records. (26 March 1953; WR1, p. 97)
\end{quote}

and also, for the first time in extant writing, coined one of his snappy slogans to describe his project, a ``universe of particles'' (WR1, p. 108). Indeed, though still bogged down in the attempt at formulating a liaison theory of gravitation alone, Wheeler always had in the back of his mind the further goal of combining this with electromagnetism and thereby achieving Einstein's goal of a unified theory (though without fields), and ultimately push on to include also the intrinsic properties of particles, such as spin:

\begin{quote}
Don't feel discouraged about how much will still remain to do after expressing mere gravitation theory in liaison form. Should serve as guide in trying to put combined gravitation-electromagnetic theory in liaison form, and in later trying to put everything in neutrino language... (WR1, p. 113; 8 April 1953)
 \end{quote}
 
The term ``neutrino'' appears here for the first time prominently in Wheeler's relativity notebook. Its significance for Wheeler is somewhat hard to grasp, as it can imply two distinct things: It appears as the barest possible point particle, carrying no charge or mass (only spin, possibly), or it can appear as a spinor field, the elementary carrier of spin and associated with the weak nuclear interaction, a reading that goes back to Wheeler's 1945 American Philosophical Society talk, where he referred to the neutrino as a ``field of interaction.'' This should be kept in mind in the following. What both notions have in common is that the neutrino is associated with the introduction of spin into the theory, which also appears to be the role in which it is invoked here. As the hope of recasting general relativity in liaison form faded, the fleshing out of the world line picture, i.e., the construction of a more sophisticated model of matter that would also include intrinsic properties such as spin, moved to the center of Wheeler's thinking.
 
 \section{Particles as Singularities in the Field}
 \label{sec:singularities}
 
 At some time in the spring of 1953, a shift began to occur in Wheeler's research agenda. Despite the day-to-day evidence we have from his notebooks, it is hard to date it exactly. It was rather a gradual shift, even though Wheeler's later use of religious metaphors to describe this tradition might rather imply an instantaneous conversion:
 
\begin{quote}
The idea of action at a distance I gave up, not because the action and the distance was complicated, but because the particle was complicated. It was just the wrong basic starting point for the description of physics, to think of a particle. Pair theory made clear, and renormalization theory, that what one thought was an electron was really an infinite number of pairs of positive and negative electrons indeterminate in number and that the whole of space is filled with pairs. [...] \emph{And of course nobody gets religion like a reformed drunkard.} As I've often said about this subject, the fanaticism, if you would like to call it that, with which I pursued the opposite approach---that it's a pure field theory explanation of nature that one ought to work at---comes from having worked so hard at a pure particle explanation of what one sees. (LW, emphasis by us)
\end{quote}

Interestingly the reasons that Wheeler gives for abandoning the particle approach (in particular the rise of renormalization theory) may well have been essential for his abandoning of the ``everything as electrons'' program, but played no role for his assessment of action-at-a-distance gravity, which, as we have seen, he was pursuing well into the 1950s. And his shift to field theory did not initially involve thinking of the particle as something ``complicated.'' Rather, he merely shifted from thinking of the particle world lines as the primary elements of the theory to thinking of them as secondary, derived objects, as singular lines in the field, whose equations of motion could be derived from the (vacuum) field equations simply by requiring consistent boundary conditions. This program goes back to the 1920s \citep{einstein_1927_allgemeine}\footnote{See \citep{havas_1989_the-early} and \citep{lehmkuhl_2017_general}.}. Wheeler focused primarily on the approach by Leopold Infeld, which was first formulated in a paper by Einstein, Infeld (then Einstein's assistant) and Banesh Hoffmann \citep{einstein_1938_the-gravitational} and consequently goes by the name of EIH. It remained a major focus of Infeld's research all through the 1940s.

Wheeler had been interested in EIH early on and, in \citep{infeld_1949_on-the-motion}, he is in fact credited with pointing out the fact that the EIH program has only a trivial (Minkowski) zero-mass limit, and that consequently a separate proof is needed in order to show that test particles follow geodesics in a non-trivial background field. The first reference to a paper by Infeld in Wheeler's notebook, however, appears only on 14 April 1953 (WR1, p. 125), several days after his last attempt to construct a liaison action (using the divergence of liaison lines in a small volume element) had ended inconclusively. Already in that attempt he had had to assume a pre-exisiting (though not necessarily metric) space in which to place the volume element. Wheeler was thus setting aside his ambitious goal of reconstructing space and time entirely from the world lines and liaisons, hoping that ``that deduction will come later'' (WR1, p. 113). Turning to the Einstein-Infeld-Hoffmann approach was a further step in this direction. After an intense study of Lichnerowicz's formulation of general relativity as initial value problem,\footnote{He had been pointed to these mathematical works by Arthur Wightman; WR1, p.121.} which he hoped to combine with the EIH approach (the notes carry the header ``Geodesics from Field Eqns \emph{or} Initial Conditions on Field Eqns''), he formulated, on the last pages of his first relativity notebook, a new research program on 1 May (WR1, p. 150).

Before we turn to this research program, we should briefly discuss the attraction of the EIH approach. For it is quite striking that only a few years earlier the EIH approach had been adopted as the basis for another attempt at a theory of everything, Peter Bergmann's construction of a theory of quantum gravity.\footnote{For more details, see \citep{blum_2016_quantum}.} Bergmann's hope had been that by transferring the EIH approach to quantum theory, the equations of the quantum mechanics for point particles might follow from the quantum field theory of general relativity in a similar manner as the classical equations of motion for point particles could be derived from the classical field theory. Even though Bergmann and Wheeler were pursuing quite different approaches, their common interest in EIH can be explained rather simply: EIH held the promise that general relativity might have something to contribute to the microphysics of particles. And for Wheeler, who had now been trying unsuccessfully to reconstruct general relativity from microscopic particle trajectories for quite some time, this prospect, which at the same time let him keep the central notion of the world line, was naturally very interesting.

For Einstein, the representation of matter particles as singular world lines in EIH had not been intended as final. It was a place holder for an ultimate (field theoretic) description of matter, no better (but also no worse) than the energy-momentum tensor on the right-hand side of the Einstein equations.\footnote{This assessment is based on \citep{lehmkuhl_2017_general}.} For Wheeler, on the other hand, coming from the pure world line approach, singular world lines appeared as a perfectly adequate description of material particles. The different status accorded to the world lines determined their assumed properties beyond mere approval or disapproval:  For Einstein the properties of the singularities could only be determined by the field equations. These did not determine the mass or the charge of the singularities, which were consequently free parameters, independently choosable for each individual singularity; much to Einstein's dismay, it should be added, as he hoped that the final theory would be able to explain why only two different masses (electron and proton) occur for the elementary particles \citep{einstein_1935_the-particle}. For Wheeler, in contrast, the world lines were still entities in and of themselves, and the default assumption (at least in Wheeler's `everything as electrons' tradition) was that they would be identical:

\begin{quote}
[T]here is no place for the $e/m$ of a particle to enter, and all particles should have the same $e/m$. (WR1, p. 150)
\end{quote}

This presented challenges of its own, since there was of course more than one type of particle in the world. Wheeler reported that his physicist colleague Hartland Snyder ``was inclined to pooh-pooh it all [on] acc't of existence of mesons, etc., in the world.'' (WR1, p. 150). Still, Wheeler was optimistic and had some ideas on how to produce a larger variety of particles with just one type of world line: Anti-particles were to be explained as world lines with the opposite orientation in time (an idea he had proposed to Feynman already a decade earlier); and he hoped to include spin in the picture by somehow taking into account the duality introduced by the two-sheeted Einstein-Rosen metric:

\begin{quote}
Their [Einstein and Rosen's] bridge idea is most intriguing -- two sheets of $g$ meeting at each singularity, get neutrino? (WR1, p. 151)
\end{quote}

On 13 May 1953, Wheeler then took his new idea of combining a (ideally unified, i.e., gravitational and electromagnetic) field theory with particles explicitly described as singular world lines to Einstein himself, when he visited him in his house on Mercer Street together with his entire relativity class. Ten years after his first discussion on AAD gravity with Wheeler, Einstein's reaction appears to have been mixed. As opposed to most of the others that Wheeler had spoken to, ``Einstein agreed [the] universe had to be closed to make [Mach's] principle valid'' (p. 11 of Wheeler's Notebook Relativity 2, henceforth referred to as WR2), but believed this to be merely a necessary but not a sufficient condition.\footnote{According to the recollections of Wheeler's student Marcel Wellner, Einstein had apparently not thought about Mach's principle in a long time \citep{wheeler_1979_mercer} when it came up during the visit of Wheeler's class. But less than a year after that visit, Einstein was asked about the matter again, by Felix Pirani. Einstein expressed his surprise at the renewed interest, opening his letter of 2 February 1954 (Einstein Papers, Jerusalem) with the words: ``There is a lot of talk about Mach's principle." By that time, apparently having rethought the matter following the meeting with Wheeler and his students, Einstein had convinced himself that the principle was obsolete, telling Pirani: ``In my opinion, one should not speak of Mach's principle at all any more.''}

Einstein's reactions to the specifics of Wheeler's research plan were even more lukewarm: He declared that he ``was not interested in singularities'' (WR2, p. 11) and that the idea expressed of ``connecting [an Einstein-Rosen bridge] up with spin of electron, neutrino is no good.''\footnote{Arthur Komar offered a more specific account of Einstein's dismissal of Einstein-Rosen bridges, recalling: ``John Wheeler asked him about the Einstein-Rosen bridge. Why had he first introduced it and then dropped it again? Einstein answered that he had initially believed that the bridge connects two almost plane surfaces in a unique manner. When he, however, discovered that they did not have a unique structure, the bridge seemed to him to be too cumbersome, unattractive, and ambiguous.'' (\emph{John Wheeler fragte ihn \"uber die Einstein-Rosen-Br\"ucke. Warum habe er sie zun\"achst eingef\"uhrt und dann wieder fallengelassen? Einstein antwortete, dass er zun\"achst glaubte, die Br\"ucke verbinde zwei fast ebene Fl\"achen in eindeutiger Weise. Als er jedoch entdeckte, dass sie keine eindeutige Struktur war, schien ihm die Br\"ucke zu schwerf\"allig, unattraktiv und vieldeutig.}) These still rather vague recollections might be of Einstein referring to the fact that he had hoped that multi-bridge solutions to the Einstein equations might be so constrained as to enforce equal masses for the individual bridges, thereby addressing the problem discussed earlier of explaining why only a few different mass values for elementary particles were observed. He ultimately appears to have concluded that no such constraints would arise, as stated in a letter to Richard Tolman of 23 May 1935 (Einstein Papers): ``One does not see why the ponderable and electric masses cannot be arbitrarily large or different, when several are present.'' Many thanks to Dennis Lehmkuhl for discussions on the Einstein-Rosen paper and for making this letter available to us.} (WR2, p. 11) Wheeler's general relativity class ended two weeks later with a final exam on 28 May. His interest in general relativity was unbroken and his notebook contains notes on cosmology, gravitational radiation, and long passages in French copied from Lichnerowicz's 1948 lecture notes ``G\'eom\`etrie diff\'erentielle et topologie''  before having to return them to the library (WR2, pp. 29-34). But for two months after the visit to Einstein, the notebook contains nothing new on Wheeler's foundational ideas and the question of how to turn singular world lines in general relativity into full-fledged particles. Wheeler did take Einstein's negative remarks with a grain of salt, in particular attributing Einstein's negative attitude toward singularities to the fact that recent work by Infeld had shown that applying the EIH method to Einstein's unified field theory did not return the correct equations of motion, i.e., the Lorentz force law in curved space-time \citep{infeld_1950_the-new}. But it was his preparations to give a talk at the International Conference of Theoretical Physics in Japan, to be held in September 1953, that gave Wheeler a new impulse.

\section{Daring Conservatism and the Field Program}
\label{sec:fields}

Wheeler ended up giving three talks in Japan: two rather technical ones on the origin of cosmic rays \citep{wheeler_1954_the-origin} and on collective models for nuclei \citep{wheeler_1954_collective}, published in the conference proceedings; and one more programmatic talk, which he held on 10 September 1953, before the conference, at the Physical Society of Japan and which was only published in Japanese translation\footnote{According to the notebook (Wheeler Papers) that Wheeler kept during his stay in Japan, the translation was done by Takahiko Yamanouchi; Japan Notebook p. 51.}  in the Proceedings of the Society. We provide a retranslation into English of this talk (the original manuscript and recording are lost) in the appendix. It is this talk which is of central importance to our story, and it is this talk that one finds Wheeler preparing in his notebook on 18 July 1953 under the heading: ``Philosophy of approach to elementary particle problem''. From the start, Wheeler was very eager to establish a clear connection to Japan in his talk, noting in the margins: ``Each one of us finds himself reflected in the countries he visits.'' But he also took the opportunity to reflect on his overall methodology. We have seen several times Wheeler's predilection for taking existing theories and using and extrapolating them outside their established domain of applicability, the paradigmatic example being his attempts to explain the nuclear forces electromagnetically. In the notes for the Tokyo talk, this methodology, which he would later characterize as ``daring conservatism'' is now, for the first time, made explicit as the ``Tokyo Program'':

\begin{quote}
Proposed Tokyo program: Be as conservative as possible about introducing new elements into description. Make basics as clear \& simple as possible. Is only the consequences that are complicated: ice; elem. particles; meteorology; geology. [...] Strengths of this approach. Its weaknesses. Einstein's May '53 remark to JAW: `The Lord may have made the universe with five fields. I don't think so. But if he did, I am not interested in the universe.' Quote as a Princeton physicist, nameless. An extreme attitude, not fully open minded. Surely much good.
\end{quote}

As he outlined his guiding methodology explicitly for the first time, Wheeler was clearly becoming excited, referring to himself in the margins as ``Tokyo Wheeler'', drawing an admittedly somewhat bizarre analogy between his new ideas on elementary particle physics and the demoralizing propaganda spread to the American troops by ``Tokyo Rose'' (Iva Toguri), host of the WWII Japanese English-language radio show ``The Zero Hour''.

At this point Wheeler's notes shift away from a lecture sketch to an inner monologue about the foundations of his research program:

\begin{quote}
Evidently have in mind something more fundamental. Out with it! Desert island philosophy: imagine selves cast up on Wake Island with library of all theory \& exp[erimen]t up to now, to solve elem. particle problem -- What to use as starting points? -- Others not ambitious enough? Go whole hog now!
\end{quote}

What follows is a long list of elements (of existing theory) that might be of importance in his attempts at crafting a theory of elementary particles. We explicitly see Wheeler assessing the potential of existing theory, in particular general relativity. The list contains familiar tropes (action at a distance -- point 2; Mach's principle -- point 7), but also some novel elements, indicating how Wheeler was reordering his vision of how to think of elementary particles. The central new element is an emphasis on fundamental masslessness, the vision of a theory without intrinsic mass parameters that would ideally include``\emph{no natural constants}. Nothing but $e$, $\pi$, etc.'' (point 1).\footnote{In a manuscript entitled ``The Zero Rest Mass Fundamental Field Hypothesis'' (WR, p. 101), which we shall discuss later in more detail, Wheeler ascribes this vision of a theory with no free parameters to Einstein. We have not been able to find relevant statements in Einstein's work.} Where then was mass to come from? In point 8 of his list, Wheeler remarked on the ``Electromagnetic origin of mass and the self energy story'', jotting down the first order radiative corrections to the electron mass, as first derived by Weisskopf in 1939. These terms were, in modern theory, simply absorbed in a renormalization of the electron mass, ultimately implying total agnosticism about the origin of mass. But in view of the proposed masslessness of the fundamental pointlike particles, Wheeler was highlighting the electromagnetic origin of mass, advocating (point 11) that one ``should apply electrodynamics to very small distances''.\footnote{Wheeler here also invoked, for the first time, Bohr as the godfather of daring conservatism, because Bohr had applied ``electrostatics to very small distances'' in his atomic model.} With mass externalized from the point-like singular particles to the surrounding field, Wheeler could consider all particles as composite (point 13), as ``structures held together by radiative, electrodynamic and gravitational forces.'' (point 14) Wheeler's new vision thus really amalgamated all existing theory by proposing particles with a singular point-like core and field-generated structure. 

This new focus on masslessness temporarily moved the neutrino to the center of Wheeler's theorizing as he emphasized the ``importance of the \emph{neutrino} in the scheme of things'' (point 3). We again encounter the ambiguity in the conceptualization of the neutrino: At one point it appears as the fundamental point-like entity, with the electron to be thought of as a ``neutrino with a charge loaded on its back''. At other times, it clearly appears as a field-like entity, possibly arising through ``spinorization'' of the metric of general relativity, that is taking the ``square root'' of the (vacuum) Einstein equations in a manner analogous to that which generates the Dirac from the Klein-Gordon equation. While Wheeler saw this as a major challenge, he was rather optimistic, remarking: ``Spinorize, fit all together, and listen for the harmony.''

Wheeler's novel emphasis on neutrinos was apparently also fueled by first results of the efforts by Frederick Reines and Clyde Cowan to directly detect these elusive particles. At this time, in the summer of 1953, Reines and Cowan were performing first background checks with their liquid scintillator detector at the nuclear reactor in Hanford, WA. They had found a source-independent background, which they thought might be due to ``natural neutrinos'' (what one might call cosmic relic neutrinos in big bang cosmology). Wheeler was aware of these results, referring in his notebook to ``Reines-Cowan radiation'' when emphasizing the importance of the neutrino. Wheeler was briefly envisioning the neutrino not only as the fundamental constituent of all particles, but also as the prime component of the energy density of the universe, and his notes of 6 August 1953 show him studying the Friedmann equations in a neutrino-dominated universe.\footnote{No correspondence between Wheeler and Reines or Cowan from 1953 is extant, but Reines in turn was clearly aware of Wheeler's contemporaneous elevation of the neutrino to central stage. In his Nobel lecture, Reines makes an inside joke, remarking without mentioning Wheeler: ``While we were engaged in this background test, some theorists were rumored to be constructing a world made predominantly of neutrinos!''} These calculations were interrupted by a phone call from Reines, informing Wheeler that they had identified their source-independent background as due to nuclear capture of cosmic ray muons.

Still, the neutrino kept an important role, also in the talk that Wheeler eventually held in September 1953 in Tokyo. The talk is set up as a dialogue between Wheeler and two figures from Japanese history, Saigo Takamori and Sugawara no Michizane.  Saigo, an important 19th Century Samurai, is given the role of the daring modernizer and presents the current state of the art in particle physics, the discovery of new particles at accelerators, and the meson theory of nuclear interaction. Sugawara no Michizane, a Ninth Century scholar and poet, is given the role of reflective traditionalist, who presents Wheelers Tokyo program, though the program is not actually named in the talk. It is merely characterized as ``the principle, which is the basis of the scientific method, of not introducing a new hypothesis until it is clearly and undoubtedly necessary.''

Sugawara begins by lauding general relativity as a model field theory: On the one hand, there
is the point we have already discussed extensively, that it allows for the integration and the
derivation of the equations of motion of point particles. But more importantly, general relativity, viewed as Einstein's formalization of Mach's principle, was supposed to provide an
account how a field theory (or more generally an interaction, which could also be a theory
of action at a distance) could generate mass in a massless theory, or rather inertial mass in a
theory without inertial mass. The argument as presented in the talk (or at least as presented
in the Japanese translation) is somewhat elliptic. It is formulated not in terms of general relativity, but in terms of an AAD theory. Clearly, such an AAD theory could not be equivalent
to GR; we know that Wheeler had been searching for such an AAD formulation of GR for several years, but had not been able to construct one. Instead, the AAD theory he used in the
Tokyo talk was a slight modification of Newtonian theory, where the usual Coulomb field is supplemented by a second field that falls off only as $1/r$ and thus dominates at long
distances. Wheeler gives this field explicitly as $G m_g a/c^2 r$, where $m_g$ is the particle's \emph{gravitational} mass and $a$ is its acceleration. This expression is analogous to the long-distance Li\'enard-Wiechert field of an accelerating charge in electrodynamics, and since it was not to be expected that the analogy between electrodynamics and gravity would be that perfect, Wheeler/Sugawara put the expression in scare quotes. With the long-distance interaction established, Wheeler then introduced what he called the "whole idea of gravity theory", namely that the total gravitational force on a particle is zero: a particle subject only to gravity is not moved by forces, but by the curvature of spacetime. In the modified Newtonian AAD theory the same idea is appropriate to express the expectation that inertia is provided by interaction and an "intrinsic inertia" term ($m_{\mathrm{inert}}a$) is absent from the equation of motion:

\begin{equation}
\frac{G m_1 m_2}{r^2} - \sum_k \frac{G m_1 a m_k}{c^2 r_k} = 0
\end{equation}

The equation's second term can instead be understood as the reaction on mass $m_1$ to the force that $m_1$'s acceleration exerts on the masses $m_k$ through the new, long-distance, Li\'enard-Wiechert-type interaction. One gets the usual (unmodified) Newtonian equation of motion for $m_1$ in gravitational interaction with $m_2$ (with $m_{\mathrm{inert}} = m_g$), under the condition that

\begin{equation}
\label{eq:mach}
\frac{G}{c^2} \sum_k \frac{m_k}{r_k} = 1
\end{equation}

where the sum extends over all of the distant masses $m_k$ which are at distances $r_k$ from the mass $m_1$.

It is appropriate at this point to point out the intimate relation between Wheeler's argument and a sketch of the origin of inertial mass published by Dennis \citet{sciama_1953_on-the-origin} just a few months before Wheeler's talk. Sciama's argument was field-theoretical, but also built on the electromagnetic analogy, explicitly employed vector fields obeying the Maxwell equations as gravitational fields and obtaining long-distance Li\'enard-Wiechert potentials that correspond to Wheeler's long-distance force. Sciama also introduced an analogous principle to Wheelers ``whole point'', which in his field-theory language reads that ``the total gravitational field at the body arising from all other matter in the universe is zero'', but Sciama explicitly labels this as a postulate and specifies that it holds in that body's rest frame. In this rest frame, the whole exterior universe is moving with acceleration $-a$, and the total field from the distant matter (the $1/r$ term) should exactly cancel the short-distance gravitational field (the $1/r^2$ term) of the particle with mass $m_2$. Rewriting this equation of cancellation, Sciama gets the usual Newtonian force law for the gravitational interaction between the masses $m_1$ and $m_2$  under the same condition as Wheeler (Equation \ref{eq:mach}) obtained in field-theoretical terms (Equation 6/7 of Sciama). It is unclear whether Wheeler knew of Sciama's argument and merely rephrased it
in AAD terms, or whether he had found it independently in his attempts at constructing an AAD version of gravity, building on an AAD formulation of electrodynamics. Both stories seem plausible, and if Wheeler really did not mention Sciama in his talk (and this is not just an omission of the transcription that was then translated into Japanese) the second one
seems the more likely. Wheeler did eventually learn of Sciama's paper, as he jotted the reference down on the last page of his second relativity notebook (which covers the period up to April
1954), but since this last page appears to have served as a general place to note miscellaneous references, it is impossible to date. In any case, the Machian argument in the Japan talk was merely to serve as a proof of principle how mass might arise in a theory in which it is not a primary attribute of matter.

A similar proof of principle was given for the electrodynamic generation of mass through the radiative corrections calculated by Weisskopf, which we have already mentioned above. Wheeler's treatment in the Tokyo talk is somewhat problematic. Following Weisskopf, he (or rather Sugawara) gave the radiative correction $\delta_m$ to the electron mass as

\begin{equation}
\label{eq:weisskopf}
\frac{\delta m}{m} = \frac{3}{2 \pi} \frac{e^2}{\hbar c} \ln{\frac{\lambda_{max}}{\lambda_{min}}}
\end{equation}

Leaving the question of the infrared and ultraviolet cutoffs in the logarithm aside for the moment, the parameter $m$ is here the electron's bare mass, which should be zero according to Wheeler's assumptions. Wheeler, however, takes it to be the electron's physical mass, assumes this to arise entirely from radiative corrections (i.e., from the field), and thus sets the lefthand side of the equation to 1. Today it is well established that perturbatively a massless fermion cannot gain mass from its electromagnetic interaction, precisely because the radiative corrections are always proportional to the bare mass (due to chiral symmetry). However, chiral symmetry may well be broken through non-perturbative effects, so that the general idea of a purely electromagnetic mass is not implausible. And again, Wheeler appears to merely have been floating some rough ideas for how mass might arise in a fundamentally massless theory and how one might obtain a unique value for the fine structure constant.\footnote{Here Wheeler was following in the footsteps of a number of famous physicists who had attempted to derive the value of the fine structure constant (which for a long time looked like it might be precisely 1/137) in the preceding decades. See \citep{kragh_2003_magic}.}

All of this was thus an elaboration of the program he had outlined in his preparatory notes. The conservative Tokyo Program was now personified by the measured statesman and poet who was filled with a ``love of Japanese beauty and harmony'', who wished to work only with well-established entities and theories and to introduce no free parameters, such as masses, into his considerations; though the end of the talk saw Sugawara reconciled with the audacious Samurai Saigo, already heralding the reformulation of Wheeler's program as not merely conservatism but ``daring conservatism'' several months later. The part of Wheeler's program that was most in flux, however, as witnessed not only by the Tokyo talk but also by the notebook entries of the time, was the exact role of the neutrino. While the talk clearly focussed on the field-theoretical aspect of the neutrino, it explicitly raised the question whether it was to be thought of as a massless field that joined the electromagnetic and gravitational fields in giving structure to the elementary particles, or whether it was only a derivative of the gravitational field, arising upon spinorization. 

Through his study of the literature on spin in general relativity (specifically \citep{pauli_1933_uber-die-formulierung}), and through discussions with the Princeton mathematician Oswald Veblen (30 October 1953), Wheeler reached the conclusion that the last point was true, but that this spinorization could only occur upon quantization:

\begin{quote}
My conclusion? I know that the neutrino obeys Pauli statistics, therefore cannot come into a classical theory, therefore ought to show up only after quantization, therefore I should look for the classical theory \& then quantize it a la Feynman, but with a square root, antisym, spinor character all put in at that time.
\end{quote}

The neutrino and the issue of spin, which had temporarily been at the focus of Wheeler's interest and of the Japan talk, was thus temporarily set aside and relegated to the quantum realm. This further strengthened the focus on the classical fields of electrodynamics and gravitation, which, despite the persistence of singular point particles, were doing the work. It was the fields that had the potential to clarify the question of elementary particles, that would generate masses and define equations of motion, classically and in quantum theory. Wheeler's main focus was thus now on Einstein-Maxwell theory, a classical field theory that would, at least after quantization, give a full account of the physics of elementary particles:

\begin{quote}
If $\nu$ is somehow contained in em+grav., and if we are right saying that only fields of zero mass count (no meson fields, etc.), and if we have the \emph{right} theory of em+grav., and if Feynman procedure [path integral quantization] is legitimate for such fields, then \emph{here's where we start}.
\end{quote}

In Einstein-Maxwell theory, the electromagnetic and gravitational fields appear as separate entities and are simply minimally coupled. Wheeler referred to it as the ``un-unified field theory''. The contrast with the unified field theory program of Einstein and others was clear and indeed these were to be viewed at the time as legitimate competitors of Einstein-Maxwell theory as classical descriptions of electrodynamics and gravitation. Wheeler thus felt the need to consider their merits, before further pursuing his program. 

How now to judge these merits? Einstein's unified field theory \citep{einstein_1950_the-meaning} was out, because, as we have already mentioned, one could not EIH-derive the Lorentz force from it. But Wheeler's student Arthur Komar (23 October) had pointed him to an alternative unified theory that gave, through the EIH method, the correct equations of motion, i.e., including the Lorentz force. This was the unified field theory of Behram \citet{kursunoglu_1952_einsteins}. There was, however, a different problem with Kursunoglu's approach for Wheeler: It relied on the introduction of a fundamental length, i.e., a dimensionful parameter into the theory, which was of course in strict opposition to Wheeler's program of having no natural constants. Wheeler asserted that ``conservative me'' (30 October) had to try out what would happen in Kursunoglu's theory when one let the fundamental length go to zero: Would one still have a unified field theory or would one merely obtain general relativity without electrodynamics? The above quote thus continues:

\begin{quote}
Only one question \emph{before we start} --- what about so-called unified field theory? Einstein's variety no good. Therefore try Kursunoglu's variety --- in case where his $p$ [inverse of fundamental length] is set equal to $\infty$ --- just to test whether we have any \emph{conservative} alternative to what we are doing.
\end{quote}

Wheeler was thus now explicitly using conservatism (in the sense of no natural constants) as a criterion for theory selection. On 1 November, he came to the conclusion that Kursunoglu's theory, in the limit where the fundamental length goes to zero, merely reproduced Einstein-Maxwell theory. His assessment of unified field theory thus ended with a ``bronze plaque'' in his notebook, reading: ``Unified Field Theories died here'' and a letter to Kursunoglu, on 3 November, in which Wheeler wrote:

\begin{quote}
I am writing to ask if a conservative physicist who wants to deal with gravitation and electromagnetism within the framework of general relativity has nowadays any acceptable choice but to use as action the expression [action of Einstein-Maxwell theory]. By ``conservative'' I mean unwilling to  introduce new ideas, new concepts, and particularly unwilling to introduce any quantity with the character of a fundamental length except as called for by inescapable evidence.\\
Will not one who adopts the conservative point of view, as just defined, have to abandon unified field theory as it stands at present?
\end{quote}

Wheeler had thus firmly convinced himself that the theory he needed to quantize was the conservative, minimal Einstein-Maxwell theory; he had found the new focus of his research in an attempt to quantize gravity, minimally coupled to electrodynamics. While quantum gravity nowadays, with all of the technical and conceptual difficulties it entails all too clear, hardly seems a conservative endeavor, to Wheeler it certainly seemed as such; it was based merely on a combination of the well-established principles of general relativity, Maxwell electrodynamics, and quantum theory. After 20 years of private speculations, he felt he was now ready to publicly elaborate on his vision for the foundation of physics, a vision that was built on general relativity, a theory that was not only coherent and well-established, but also, through its unique features, such as Mach's principle and the EIH determination of equations of motion, had the potential to resolve the great open questions of microscopic physics. On 4 November 1953, we thus find in Wheeler's notebook ``Points for proposed article `Elementary particles from Massless Fields --- An Assessment.''

Around this time, Wheeler suddenly appears to have remembered a central point, which indeed was absent at least from his notebook entries for a while: the point particles. For November 8, we find the following short entry:

\begin{quote}
The big question\\
Let's forget about electromagnetism for present. In quantum transcription of the pure gravitation theory with the variation principle based on $\psi = \sum e^{\frac{i c^3}{16 \pi G \hbar} \int \int \int \int R \sqrt{-g} dx^1 dx^2 dx^3 dx^4}$ [i.e., the path integral for the Hilbert action] how do we take into account the existence of singularities?
\end{quote}

The singular world lines had, over the course of the year 1953, been transformed from the central element of the theory into a problematic embarrassment in the promising program of quantizing general relativity. Like Bergmann several years earlier, Wheeler realized that quantization and point singularities in the field did not really mesh. Bergmann had resigned himself to studying pure general relativity, but this was hardly an option for Wheeler who was after all trying to solve the problem of elementary particles. And indeed, the fields in Wheeler's approach were still mainly meant to provide services to the point particles: give them mass, define their equations of motion. When he met with Einstein once more, in the morning of 13 November 1953, Einstein asked (WR2, p. 83): ``What about matter term in Lagrangian''  to which Wheeler replied that ``matter was to originate from singularities.'' However, when Wheeler then went on to explain Feynman quantization to Einstein, he remarked that in this setup ``the singularities in field get eliminated, never have to be talked about.'' This seems to be in reference to the assumption that singular field configurations would have measure zero in the path integral, which seems like a problem for describing matter by singularities, but is of course a good thing when talking about pathological singularities.\footnote{Indeed, Einstein appears to have been impressed. While first remarking that he ``abhorred'' the idea of first constructing the classical field theory and then quantizing it, he then conceded (according to Wheeler's notes) that ``it was the first time he had ever heard describe a way that [quantum theory] might get through, found it very attractive.''}

But an even more severe difficulty with the singular point particle notion lay in its relation to the field concept. EIH determination of the equations of motion, of course, offered the prospect of reconciling  the notions of field and point particle; this fact had originally led Wheeler to reintroduce fields into his worldview and endorse a dualistic ontology. As soon, however, as mass generation through the field entered into the picture, the fundamental incompatibility of point particles and local fields, which had haunted fundamental physics ever since Hendrik Lorentz had first tried to think the two together in his electron theory, again became visible. Indeed, already in his Tokyo lecture, Wheeler had been forced to introduce an ultraviolet cutoff ($\lambda_{\mathrm{min}}$ of Equation \ref{eq:weisskopf}) to make the field-generated mass finite. This essentially meant abandoning the idea of a point particle and introducing a finite size for the electron. It is important here that Wheeler (or Sugawara) had hypothesized that this finite size would be given by the gravitational (Schwarzschild) radius of the electron, and not the Planck length. So the necessary mass scale that one needed to make a length using the gravitational constant $G$ and the speed of light $c$ was provided by the mass $m$ of the electron and not by Planck's constant $h$. This clearly indicated that the cutoff was to arise not as a quantum effect, but due to the presence of the particle. By introducing the notion of field-generated masses, Wheeler had thus effectively abandoned the notion of a point particle that had been a mainstay of his research program for a long time. This was not a problem for the EIH determination of the equations of motion, as the use of point particles in that derivation could well be viewed as a mere approximation.\footnote{While \citet{wheeler_1961_geometrodynamics} would later conclude that point singularities were not a valid approximation for any reasonable model of matter (which by that time for him meant geons and wormholes), there is no indication that he (or anybody else) harbored such doubts in 1953/54, given that the concepts and in particular the conception of matter that these conclusions were based on had not been developed yet.} But it ultimately undermined Wheeler's briefly-kept hopes for a dual theory of point particles and fields and forced him to consider novel conceptions of matter.

While he spent the next weeks thinking about how to spinorize Einstein-Maxwell theory by taking the square root of the Lagrangian in the action (WR2, p. 88), the pressing question of the constitution of matter moved to the center in a working paper entitled ``The Zero Rest Mass Fundamental Field Hypothesis'' and dated 19 January 1954.\footnote{The paper is included in WR2, p. 101, as an insert. This copy is noteworthy also for some remarks in the margins in which Wheeler explicitly connects his conservative heuristic in physics with conservatism in politics, noting: ``To defend well established physical ideas as unpopular as defending well established political parties. People like to criticize. Religion the great defender.'' In this connection it appears pertinent to mention that Wheeler's conservative stance (in physics), as outlined in the Tokyo talk, was explicitly criticized by the Japanese physicist Shoichi Sakata, an outspoken Marxist \citep{staley_2004_lost}. In discussions on September 18 at the conference in Kyoto, a week after Wheeler's lecture, Sakata remarked: ``I am convinced the future theory should not be the progressive improvement of the present theory. At the Tokyo meeting Professor Wheeler pointed out that there are two methods of approaching the truth; that is Saigo Takamori's method and Sugawara Michizane's method. But in Japan Professor Tomonaga had pointed out that there are two ways, namely a non-reactionary conservative way and also a revolutionary way. This is our common sense.'' \citep[p. 34-35]{japanproceedings}} Here, Wheeler addressed the central question that any theory of extended (i.e., not pointlike) particles would have to answer. While the spatial extension of the particles avoided the issue of divergent field strengths, it brought with it a different issue, which had a long tradition going back to first attempts at a solution by Poincar\'e: the issue of stability. Given that there would be no more singular point-like cores, all that was left for constructing a particle were the electromagnetic, gravitational, and possibly neutrino fields (the ``zero rest mass fundamental fields'' of the manuscript's title), and ``an elementary particle is held together by the balance of gravitational, neutrino, and electromagnetic forces'' (p. 7 of the manuscript). But how to envision such an object? In the manuscript, Wheeler explored the possibility of comparing elementary particles with a (collapsing) star---the analogy being based on both objects (star and particle) being held together by gravitational forces.

But the big breakthrough for how to model elementary particles only occurred about a week later, when Wheeler attended the Fourth Rochester Conference on High Energy Nuclear Physics from 25-27 January 1954 \citep{rochester_1954}. It is the last one of Wheeler's breakthroughs that we shall discuss in this paper, as it finally brings us to  Wheeler's geon paper \citep{wheeler_1955_geons} and his embrace of a pure field theory, from which also the singularities representing matter had been removed. Up until now, Wheeler had mainly attempted to use the untapped potential of general relativity as it related to mass points: The ability to derive their equations of motion from the field equation, the possibility of generating mass for them from fields or interactions. In late January 1954, Wheeler seized upon a feature of general relativity, which he had hardly engaged with so far: the non-linearity of the field equations, which in principle allowed for solutions describing a localized and (meta)stable concentration of energy, an idea which had been in the back of Einstein's head for a long while. 

On his manuscript of 19 January (which was never published), Wheeler had noted that he was distributing it to a small number of physicists, including Einstein, Bohr, and Wightman. Wightman was also attending the Rochester conference, and Wheeler appears to have discussed his ideas with him there, for on 25 January 1954, we find the notebook entry (WR2, p. 96):

\begin{quote}
Ball of light held together by gravitational forces as classical model for an elementary particle = fireball = (Wightman name) Kugelblitz
\end{quote}

immediately followed by calculations for a spherically symmetric graviational potential fulfilling the vacuum Einstein-Maxwell equations (i.e., the Einstein equations with only an electromagnetic energy-momentum tensor as a source), with all of the electromagnetic energy constrained to a sphere of finite radius. Such a field configuration, which could only exist in a non-linear field theory such as Einstein-Maxwell theory and which Wheeler would soon label a ``geon'' (first found in WR2, p. 104, in an entry dated 19 February 1954), was thus the new model for elementary particles that Wheeler would pursue for the next few years. Everything point-like had been expelled from the model, in favor of a spatially extended pure zero-mass-field configuration.

There were of course many open questions to tackle, some of which Wheeler listed in the entries of the next two days (WR2, p. 100ff), such as whether such entities really existed, how to incorporate charge,\footnote{Here Wheeler already pondered the possibility of having ``outgoing lines of force [...] understood in terms of lines coming in from an `internal universe''', an idea that would later mature into his notion of a wormhole.} the still unsettled role of the neutrino and the square root of the Einstein-Maxwell Lagrangian, and the role of quantum theory and self energies,\footnote{Here Wheeler encountered some conservative resistance from Wightman, who objected to Wheeler's predilection for path integrals, arguing instead that one should ``improve \& understand present formalism'', i.e., pursue axiomatic quantum field theory. Even Feynman appears to have been doubtful about the ``general utility'' of the path integral, as he had not yet been able to properly accommodate fermions.} in particular concerning the quantization of general relativity, in which context Wheeler noted (WR2, p. 103):

\begin{quote}
Try to understand whether Gupta or anyone else really know what he's talking about on the quantization of gravitation theory, esp. the comm'n. rel'ns at small distances.
\end{quote}

But while the new geon model of elementary particles brought with it a host of unanswered questions, an entire research program as it were, just days after the Rochester conference (where he had talked on charged meson decay) Wheeler certainly felt confident enough to publicly present his new idea in New York City, where he held the annual Richtmyer Memorial Lecture of the American Association of Physics Teachers (AAPT).\footnote{\label{fn:long} The AAPT was conducting its winter meeting in parallel with the American Physical Society, which conducted its annual meeting at Columbia University from 28-30 January 1954 (Physical Review, Volume 94, pp. 742ff), so that there were also many research physicists in the audience.} This Lecture, entitled ``Fields and Particles", is the last text we shall be discussing and is, as we shall see, in many ways the sum of the development in Wheeler's thinking that we have reconstructed in this paper.\footnote{The lecture was never published, but there is an extant transcript in the Wheeler Papers, in a folder entitled ``Fields and Particles.'' The Richtmyer Lecture Memorial Award had been established in 1941 to honor Floyd Richtmyer, one of the founders of the AAPT (\url{https://www.aapt.org/Programs/awards/richtmyer.cfm}). Many of the previous lectures had ben published in the AAPT's journal, the American Journal of Physics (e.g., \citep{slater_1951_the-electron, vleck_1950_landmarks, dubridge_1949_the-effects}). Wheeler had plans to publish his lecture there as well, and the folder contains two revised versions of the original lecture transcripts, which were clearly supposed to lead up to a publication. The folder also contains some correspondence between Wheeler and Thomas Osgood, editor of the American Journal of Physics, such as a letter from Osgood of 28 January 1957, which begins: ``Here is my annual letter of inquiry about the manuscript of the paper ``Fields and Particles'' that you gave as Richtmyer Memorial Lecture during the meeting of the American Association of Physics Teachers in New York, January 28-30, 1954 It ought to be published without delay.'' Wheeler in fact cited the paper in the first footnote of the Geon paper as ``to be published''. That long footnote (a specialty of Wheeler, to which this footnote here is a sort of tribute) also contained a reference to Wheeler's Tokyo talk and ``the point of view ascribed by the author to Sugawara-no-Michizane,'' making the entire footnote rather enigmatic for the average American reader of the Physical Review.} 

The Richtmyer Lecture began with Wheeler's most explicit elaboration of his conservative methodology, which he now labelled ``daring conservatism'' and couched in religious terms, citing the apostle Paul:

\begin{quote}
``Whatsoever things are true, whatsoever things are honored, whatsoever things are judged, whatsoever things are pure, whatsoever things are lovely, whatsoever things are of good repute. If there be any virtue and if there be any praise, think on these things.''\footnote{This passage is from Philippians 4:8, where it reads ``honest'' instead of ``honored'', ``just'' instead of ``judged'', and ``report'' instead of ``repute.'' We have given the quote as it appears in the lecture transcript, and it is to be assumed that the transcriber simply misheard these three words. Wheeler corrected all three in the later manuscripts of the Richtmyer Lecture mentioned in Footnote \ref{fn:long}.} Following these words of Paul, I would like to dedicate this occasion [...] to an appreciation of the great truth of physics in the saying that from them we will receive guidance in this elementary particle problem beyond anything that we now imagine.
\end{quote}

 Wheeler then went on to highlight the role of general relativity among the ``already well established ideas'' of physics on which the conservative physicist should build by daringly ``following out [its] consequences'' to the ``utter most extreme.'' He then went on to outline the great potential (``exciting new possibilities'') of general relativity both ``in the realm of what might be called astrophysics'' and for the ``elementary particle problem'', introducing his geon\footnote{Then still referred to as a ``Kugelblitz'' or, in the words of the person who transcribed the lecture, ``cugoflix''.} idea to the world and presenting it as a new research program:
 
 \begin{quote}
 In my view following out the philosophy of the conservative daring [sic], it's an inescapable obligation of our present-day physics to continue the investigation of these objects and to see what boundary line if any separates them from the elementary particle problem. The full investigation of both electromagnetism and gravitation of course has to take place within the frame work of quantum theory.
 \end{quote}
 
Wheeler had thus publicly outlined his new research program in general relativity, which consisted of studying stable, localized solutions of the Einstein-Maxwell equations, their modification through quantum theory and their relation to elementary particles, as well as the inclusion of further elements into this picture, such as charge and the neutrino/spin. Wheeler's transition to a full-blown ``relativist'' was completed, and the research program outlined in the Richtmyer lecture would occupy him and his graduate students for years to come. So fruitful was this approach that Princeton and the Wheeler School, despite being the youngest of the relativity centers soon to be connected in the Renaissance, became one of the central hubs of that process.

\section{Conclusions}
\label{sec:conclusions}

In this paper we have reconstructed John Wheeler's turn to general relativity in the years ca. 1941-1954 and how it was driven by what we have called the untapped potential of general relativity, thereby corroborating and filling with meaning the claim of \citep{blum_2015_the-reinvention} that this untapped potential was one of the motors of the renaissance of general relativity. Our reconstruction has shown that Wheeler's general methodology, ultimately branded ``daring conservatism'', precisely consisted in seeking out the potential of existing theories, rather than constructing new ones. It should, however, be added that the general notion of daring conservatism can be read in two ways, both of which Wheeler endorsed. One is to extrapolate existing theory in order to make predictions for new, unexpected phenomena and then trust those predictions, even though they are made outside the domain for which the theory has been experimentally corroborated. This view of daring conservatism is to be found in an example that Wheeler gave in the Richtmyer lecture, where he claimed that he could have predicted nuclear fission two years before its experimental discovery, had he only trusted the extreme predictions of 1930s nuclear modelling. This view also applies to the use of general relativity in making novel predictions for astrophysics.

But as we have seen, it was another reading of daring conservatism that was initially more central to Wheeler's thinking: Using existing theory not to predict novel phenomena, but to solve existing (theoretical) problems and paradoxes that one might otherwise have been tempted to solve by introducing new theories. The central issue that Wheeler came to believe general relativity had the potential to solve was what he called the ``elementary particle problem''. A precise definition of this ``problem'' is hard to come by, but it meant something along the lines of obtaining a consistent description of the internal structure of elementary particles (which originally of course implied finding a consistent theory of point-like particles without structure). The solutions that Wheeler considered to this problem were shaped by several convictions, in particular that (i) the general idea of the solution should be expressible in classical language, (ii) the solution should be monistic, or at least not gratuitously introduce various types of particles, and (iii) the solution should ideally not involve any free parameters. All of these three conditions favored Wheeler's turn to GR, which was (i) a classical theory, (ii) dealing in universal substance (space-time), (iii) involving no free parameters beside the gravitational constant (which could be set to 1 in what Wheeler would later call Planck units).

We thus see that also the further development of Wheeler's career in relativity closely paralleled the overall development, as questions relativistic astrophysics (and thus the first reading of daring conservatism) gradually supplanted (or merged with) his original foundationalist aspirations, in what Roberto Lalli, J\"{u}rgen Renn and one of the authors (AB) have called the astrophysical turn of the late renaissance \citep[p. 540f]{blum_2018_gravitational}. It turns out then that an important factor in assessing the relevance of the epistemic potential of GR in the renaissance is the question of ``potential for what?''. This is true not only with regards to what problems to solve, but also to what kind of work to generate. For we have clearly seen the strong pedagogical bent in the way in which Wheeler tackled general relativity, and the focus on problems to be solved; the general relativity that Wheeler was exploring was swarming with future PhD theses, theses in physics, that is, connecting the heretofore isolated field of general relativity to particle physics, quantum theory, and astrophysics. It was this aspect which turned Princeton from a research center among several to the home of the ``Wheeler School'' \citep{christensen_2009_john, misner_2010_john}

This brings us to a final paradox: How to explain the great impact of Wheeler's approach to general relativity, given that the various solutions to the elementary particle problem that we have discussed in this paper were all eventually viewed as misguided. Neither worldlines and liaisons nor geons are nowadays regarded as fruitful ways for thinking about the structure of particles, and also the quantization of gravity did not yield to Wheeler's simple path integral vision. Our study at least suggests an answer to this paradox: The important thing was not so much the specific manner(s) in which Wheeler tried to resolve the elementary particle problem, but rather his keen sense for which elements of general relativity would turn out to be the most fruitful. 

Looking at Wheeler's trajectory thus also provides insight into where exactly the epistemic potential of general relativity lay, namely in its unique features as a theory: the determination of the equations of motion through the field equations, the non-linearity of the field equations, and that its quantization will lead to non-trivial new physics. Conceptual studies on the role of point particles in GR could thus segue into studies on the so-called problem of motion, studies on geons into studies of exact solutions of the full Einstein equations, studies on path integral quantization would come to be regarded as important puzzle pieces in the ongoing search for a quantum theory of gravity. Here too, we observe Wheeler's trajectory closely mirroring general trends, where isolated research centers originally focusing on GR-based speculative theorizing move, in the course of the Renaissance, to the study of important conceptual questions within general relativity, relevant to the emerging community at large. The question remains to what extent Wheeler's original interests actually shaped the problems considered important in the GR community of the renaissance and beyond. But this question is beyond the scope of our study, which focused on an individual intellectual trajectory and on a conversion from particle to field theory that turned out to be far more gradual than expected. If the reader thus takes home just one fact from our story, it might be this: For a few months there, in late 1953, John Wheeler believed in both particles and fields.

\bibliography{habil}
\bibliographystyle{apalike}

\section*{Translation of John Wheeler's Tokyo Lecture}

\section*{Discussion on the Problems of Elementary Particle Theory}
\subsection*{Originally published in Japanese in \emph{Proceedings of the Physical Society of Japan} {\bf 9}, pp. 36-41 (1954)\footnote{Many thanks to the Physical Society of Japan for letting us publish this translation free of charge.}}

\subsection*{Translated by Yukari Yamauchi (University of Maryland)\footnote{The authors would also like to thank Lisa Onaga and Masato Hasegawa (both Max Planck Institue for the History of Science) for additional input. We have freely edited the translation based on our understanding of the context and the physics involved, so all mistakes should be considered ours.}}

Chairman Yamanouchi and members of the Physical Society of Japan!

I was fortunate to have a pleasant experience in Kofu before visiting here in the Tokyo region. I was deeply impressed with the energy and vision of your scientific researchers in studying the fundamental problems of physics, and at the same time, by the love of truth itself that is part of Japanese culture and philosophical tradition.

Speaking to you about the problems of elementary particles today is something quite special to me. Because I have come to a country where one of world-renowned journals on theoretical physics is published, I think that one of your groups should be speaking. But what I would like to talk about is not to discuss each special achievement, but to discuss a broad plan of research that particle physics has not attacked so far. The problem that was posed this morning - of it, we know the basic philosophy, the fundamental theory, the mathematical equations, it only remains to find the final answer. Unlike that problem, what we are facing in the field of elementary particles is to arrive at a basic principle itself. Our problem is not mathematical physics but theoretical physics. In the important mathematical and theoretical research, which we heard from Professors Mott and Slater this morning, we start from basic physical ideas that look relatively simple. I think you could see how rich and complex is the development that follows from these ideas. Unfortunately, in the field of elementary particles we still do not have the proper basic physical ideas. After thinking about how to explore this problem and about what we know from the discussion this morning, and while thinking about how to present what is not yet understood and its higher aspects, I suddenly envisaged talking to two wise men, heroes in Japanese history. One of them is Sugawara, a great statesman, a man of culture, who loved truth and beauty, and the strongest defender of liberty, who maintained his principles even risking life and love. The other one is Takamori Saigo, also a great hero, with strong personality as well as great energy and influence. In my flight of imagination it first seemed strange to be arguing about the extremely new problem of elementary particles with these two people of a long time ago and from a different era. However, soon I realized that these two people are of great interest in connection with this problem, which is attracting attention of the leading Japanese scientists. Furthermore I discovered that they had completely different opinions on this subject.

Of the two opinions, I am familiar with the one presented by Saigo. Nevertheless, he brought new ideas into the discussion. According to what he said, we already have a fundamental theory. I followed the path of his argument carefully because I am well aware of how remarkable are the advances in electron theory in the past few years, due to new methods developed partly by Tomonaga, and partly by Schwinger, Dyson, and Feynman. In this approach we try not to solve the problem of divergences, but to configure equations and ideas in such a way as to avoid talking about this blind alley of an infinite number of electrons. With this approach to the theory, it is now possible to calculate many important effects such as the microfine structure of the Hydrogen atom and radiative corrections for scattering of electrons in the Coulomb force field. I realized that Saigo Takamori thought that the meson could be treated in the same way.

We all appreciate Professor Yukawa's interesting thought of a few years ago, that the force connecting the nucleon in the nucleus is related to a new particle of intermediate mass. Thanks to this theory, new particles were actually discovered soon after. And Takamori Saigo, as well as physicists of my country and many other laboratories, thought that the same renormalization theory as that of electrons should be applied to mesons as well. In connection with this, I asked what great advance in this field is expected in the future. He interprets scattering theoretically by the formation of a complex system of one meson and one nucleon, with a fixed angular momentum. This system is virtually created and disappears, so that meson nucleon scattering is understood as a kind of resonance phenomenon. He pointed out interesting research attempting to theoretically interpret this by the idea that a resonance phenomenon occurs in the so-called ``scattering''. Many features of the problem of interaction are explained in this way, scattering of positive and negative pions on neutron and proton, and neutral pi mesons on charged pi mesons. The transformation to a phenomenological explanation by this simple idea surprises not only me but I think that we were all surprised.

\begin{figure}
  \includegraphics[width=\linewidth]{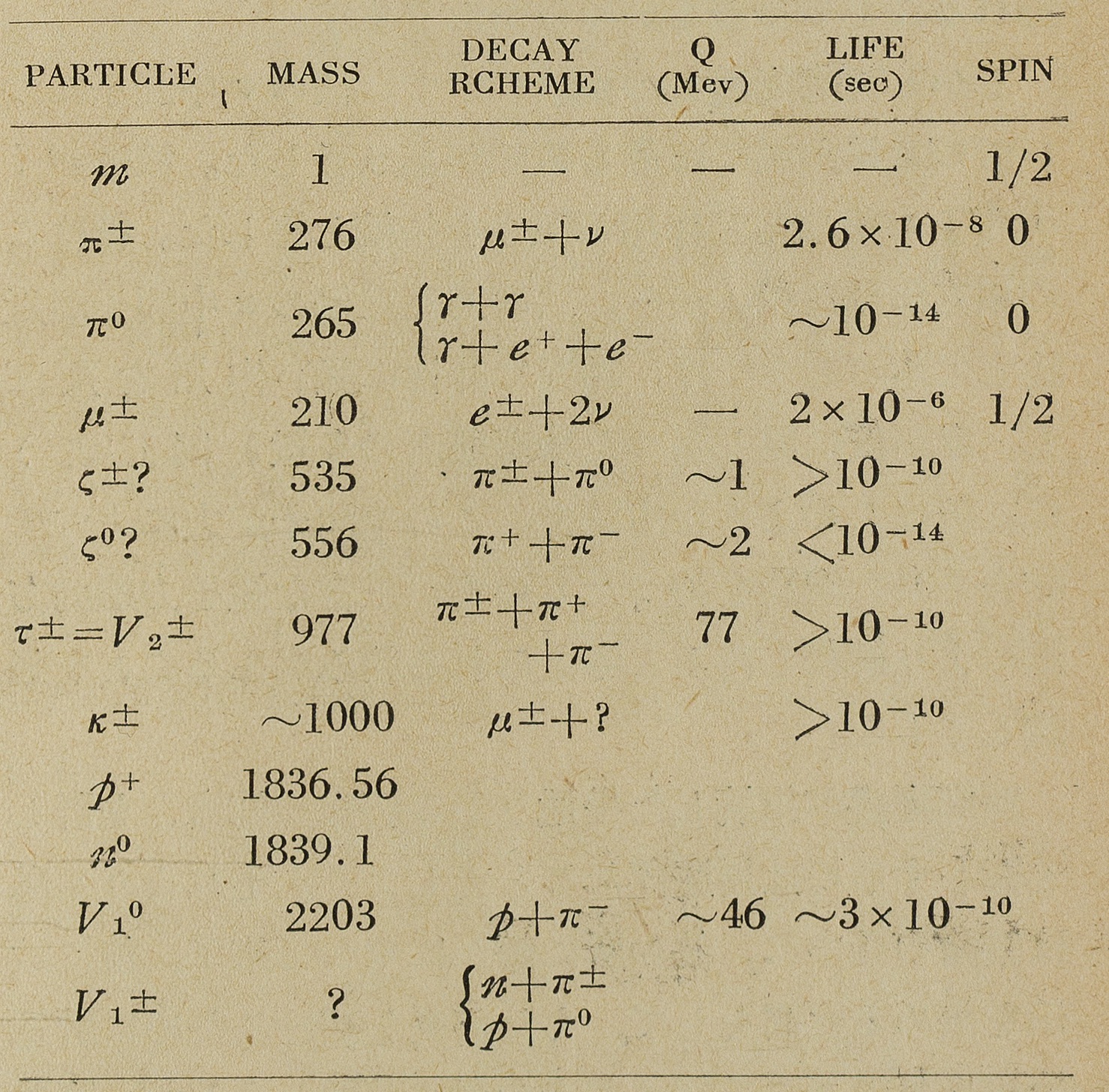}
  \caption{Table of particles discovered by 1953.}
  \label{fig:table}
\end{figure}

Next I discussed with Saigo Takamori a difficulty that appeared recently with this approach. Although it gives in this way a satisfactory explanation for experiments of scattering and conversion for energies on the order of 100 to 150 million electron volts, a recent experiment with energy reaching 1,000 million electron volts at Brookhaven seems to show that there is another resonance in the scattering and transformation cross section that does not fit the simple image of one resonance. Takamori Saigo pointed out that we have not investigated sufficiently what kind of result is to be expected for higher resonances. But I did not enter into this problem in detail, I asked, ``What is your broad general plan to get close to your problem?'' He laughed and said that as a first step to the main problem we'll discuss later, first of all let me show this table (Figure \ref{fig:table}).

This table shows what we know about elementary particles. Certainly there exist more particles that are not yet listed, Even in this table, not everything is generally accepted. Charged and neutral mesons and mu mesons are familiar and well studied, so you do not need to mention them in particular. Regarding the $\xi$ meson, there are still many questions about the existence of this particle and the details of its decay. The $\tau$ meson, however, is a most interesting particle, whose existence was first found at Mott's institute. The decay is such that this particle divides, among what is believed to be three pi mesons, the energy of 77 million electron volts, released after somewhat more than $10^{-10}$ seconds.

Let's then turn our attention to K- or $\kappa$-mesons. They have masses of around 1000, they decay into mu mesons and other, unknown, radiation. Very recently, these two particles and the $\tau$ mesons have come to be considered not different particles, but the same particle that decays differently.

I do not need to say anything about protons and neutrons. The first type of $V_0$ meson that splits up into protons and pi mesons was recently made artificially by the Brookhaven accelerator for the first time. And interesting features of the decay scheme, details of the emitted energy, and mysteries of angular correlation of the emitted particles have been found. Charged V mesons decay in the same way.

But in the discussion with Takamori Saigo, his opinion on the $\tau$ meson was the most interesting to me. He pointed out the strength of the interaction between pion mesons. As this interaction is strong, when two or three pi mesons are in close proximity, they exert a very strong effect on each other and, like many of the physicists I have already talked about, he thinks that in this way there is a hope that this $\tau$ meson can be analyzed by a normal quantum mechanical method as a structure coupled by force.

It would be foolish to talk about the manifold of studies that Saigo Takamori pointed to, about teaching or my amazement at the number of young people returning to computing machines and desks to continue the research. Prosperity brought amazing energy to all. 

Then I got an opportunity to speak with Sugawara no Michizane as Saigo got out of the discussion and was gone. Compared to Saigo, he had more lofty aspirations. He was in agreement with the study Saigo had presented, but he had different hopes for the results. He pointed out the similarity between this kind of research and the theory of superconductivity. Suppose superconductivity theory had been formed before quantum mechanics and electron theory were completed sufficiently. Since in the presence of superconductivity, the magnetic field decreases exponentially when going inside from the surface of a metal, a superconducting theory including the inherent length of exponential decay could have been made. Let us suppose that this theory were added to electromagnetic theory and electron quantum theory; that is, three fields, electromagnetic field, superconducting field and electron field should exist.

Of course it seems absurd to interpret superconductivity like this today. We have a faith that one day we can understand the mechanism of superconductivity from the theoretical principles already established. The fundamental principles and equations are simple, but how difficult it is to keep track of what results from them was best illustrated in the lecture we heard this morning. But in the case of Meson theory, is not Saigo doing exactly the same thing? Would it not be premature to introduce a new auxiliary field, the meson field? Should we not try to understand the situation we are facing without actually introducing new hypotheses and new fields? For a long time I have been impressed with the principle, which is the basis of the scientific method, of not introducing a new hypothesis until it is clearly and undoubtedly necessary. So this greatly calls for my support and respect. At the same time, however, his view was very unusual, so I naturally asked him for an explanation. He said he would like to call your attention to the nature of the theory we expect in the future. This theory says relativity should be a model. I was quite surprised to hear that Sugawara had brought up relativity, but what he said became clear soon. It is in general relativity that we have a closed theory for the first time. Of course it is incomplete, but as a theory it is self-consistent.

According to the field equations of general relativity, the reduced curvature tensor becomes 0 at each point in space outside a singular point. These equations are sufficient to describe not only the field itself, but also the motion of the particle that occurs in the field. This remarkable feature of  general relativity is consistent with the form that was desired and required for a long time as the form of a proper field theory. In addition, general relativity has another feature that is required for a basic theory. That is, it does not include natural constants or constants other than pure numbers such as 2 or $\pi$.

As I understand it, Sugawara said that it was a simple form of relativity that omits all the discussion on electromagnetic fields, the first simple relativity created to discuss mass point interactions. He pointed out that this theory is closely related to the idea of action at a distance in Newton's theory of gravity, despite being a field theory. He states that the integral representation and the differential representation of the given physical law are not dissimilar from the action-at-a-distance form and the field form of the same basic physical principle, there is no big difference between them, they are equivalent representations. He said that Mach always pointed out that space and time are the basic guiding principles of physics. According to Sugawara, the equation of general relativity is an abstraction made from this fundamental principle. Einstein made it possible to explain the inertia of matter by the interaction of a given particle with every other particle in the universe, by carrying out the program of Mach. The inertia generating force is only the gravitational analog of the radiation interaction in normal electromagnetic theory.

In this regard, the following similarities shown were the most interesting. The field generated at a distance $r$ from a given charge is $e/r^2$ at short range and $ea/c^2r$ at long range, where $a$ represents the acceleration of the moving charge. In the same manner, he emphasized that gravity is $Gm/r^2$ in the short range, but in the long range the field becomes $Gma/c^2r$, which decreases only by the first power of the distance. In this way, when one particle interacts with a very large number of other particles, one of them being nearby and the others far away, the combined force from the nearby and distant particles results in the kind of formula given below [Equation \ref{eq:five}].

The whole idea of gravity theory is that the total resultant force is zero. To be more precise, there is no inertia other than what arises from field theory. Therefore, when the particle 1 is acting on the nearby particle 2 and the other particles i are located in the very far distance of the universe, the equation of motion of particle 1 will be essentially in this form (omitting negligible terms)

\begin{equation}
\label{eq:five}
\frac{G m_1 m_2}{r^2} - \sum_i \frac{G m_1 a_1 m_i}{c^2 ``r_{i1}"} = 0
\end{equation}

Sugawara smiled at me while writing this equation and noted that the expression written here is just an approximation of the result of field theory in integral form. That is, he put quotes on the ``$1/r$'' term in the equation so that it clearly reveals that this was simply a rewrite of the true theory. He added that the second term, representing the interaction between the accelerated particle 1 with the other particles, is not exactly what we usually call the Newtonian inertial force, but explained that it is the cause of normal Newtonian inertia. Long ago, Mach brought forth the idea that inertia arises from interaction with other things in the universe, and Einstein carried out this program. According to this interpretation, that the inertial force is due to the gravitational interaction, the following sum is required to have a magnitude on the order of 1 in the universe we currently know:

\begin{equation}
\frac{G}{c^2} \sum \frac{m_k}{``r_{1k}"} \sim 1
\end{equation}

This equation seemingly includes natural constants, but Sugawara cautioned that if you use the appropriate units to measure the distances, these constants will all disappear. Thus, general relativity, although incomplete, has nothing appearing as a fundamental constant besides pure numbers such as 2 or $\pi$, one property to be required of every normal field theory.

Although this lesson is instructive, I wondered what was Sugawara's real intention about what everyone would like to hear, how this relates to elementary particle theory. To that he says, to describe the problem of elementary particles we should only use fields with a bare mass of 0 whose existence is well established. We should study whether all the particles cannot be considered to be made of such fields.

\begin{figure}
  \includegraphics[width=\linewidth]{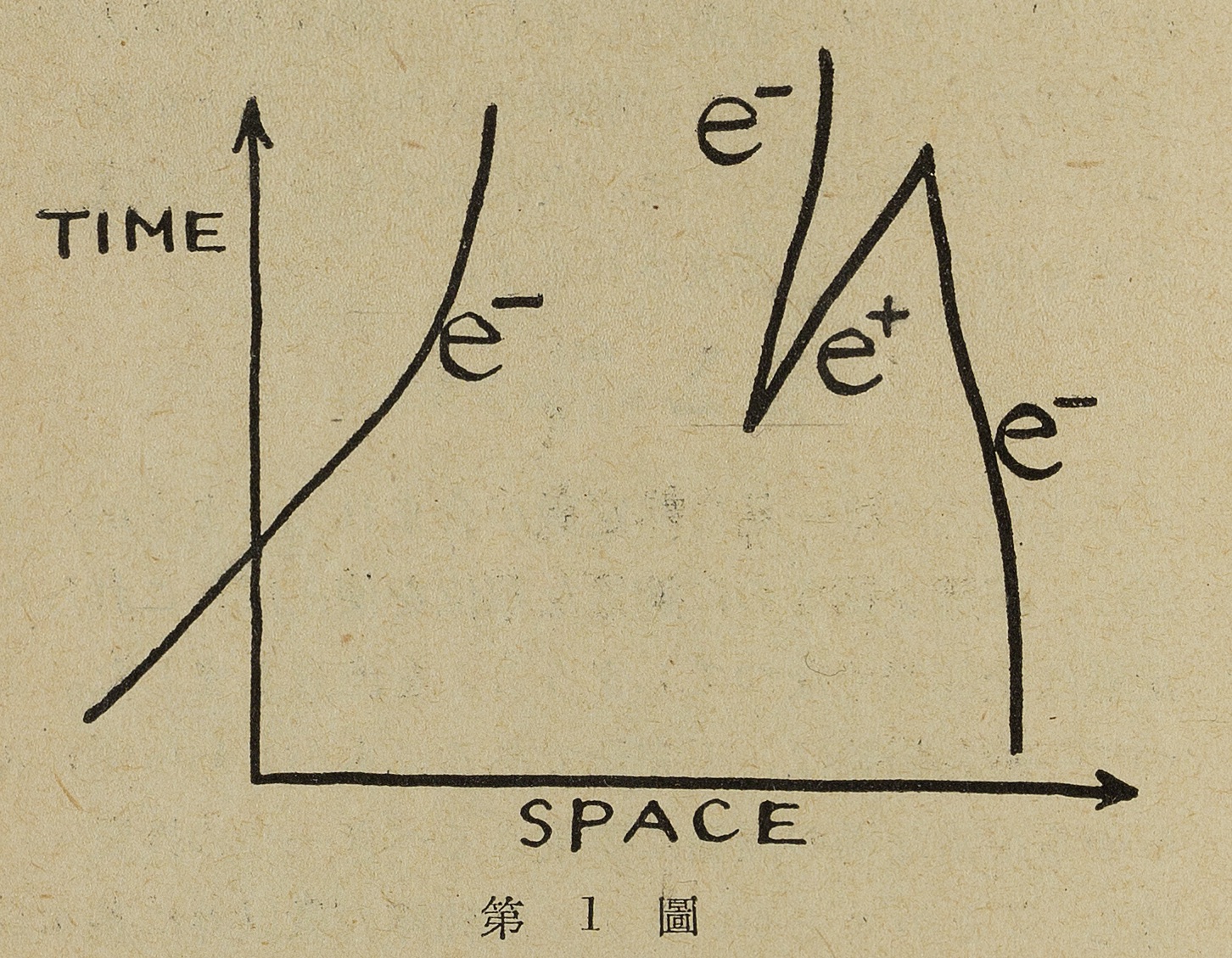}
  \caption{Positrons as electrons going backwards in time.}
  \label{fig:zitter}
\end{figure}

When the discussion had proceeded so far, I became a little worried whether such a fundamental idea should be reported at this kind of meeting. I think you and I have an obligation to consider this problem as clearly and carefully as possible. And if a person like Sugawara had something in mind, I thought that one needed to pay attention to it. In addition, it is very difficult to carry out this program. So I asked how you interpret electrons and other particles. What he said is much longer than I will mention here, but you can get the general idea in the next figure. He noted that the electrons we are now considering are described by a world line, the path of a charge in space-time. In such a diagram, it can be argued that positive electrons and negative electrons disappear at one point, and one pair of positive and negative electrons arise at other times. Alternatively, if you wish, you can use the four-dimensional format to say that the path of a charge does not advance in one direction only, in the time direction, but also turns back in time (Figure \ref{fig:zitter}).

\begin{figure}
  \includegraphics[width=\linewidth]{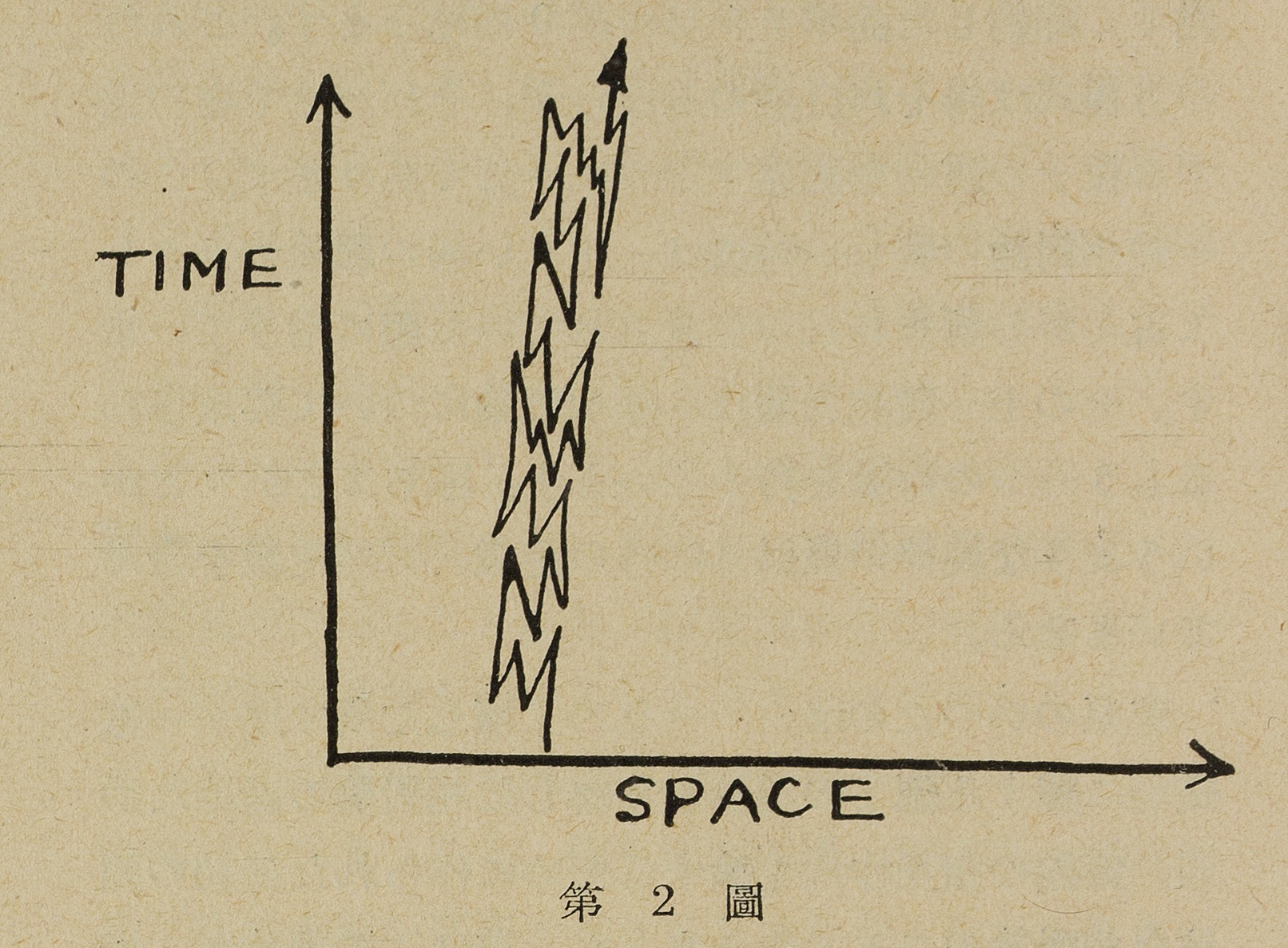}
  \caption{Electronic \emph{Zitterbewegung}.}
  \label{fig:zitter2}
\end{figure}

The next graph shows by picture what we know about the nature of electrons today. If I may use the word resolution, then if you look with weak resolving power, electrons will appear as a wide line in the space-time illustration. However, if you look at this trajectory more carefully -- speaking this way regardless of the possibility of really seeing the system -- when you look at the tracks more carefully, you can see that they go back and forth. In other words, it will be necessary to deal with the creation and destruction of many overlapping electrons (Figure \ref{fig:zitter2}).

In the immediate vicinity of a given electron, we must deal with high density fluctuations. As you can see, the divergence in electron theory arises from the fact that the interaction between this large number of virtual electron pairs becomes infinite if we examine it carefully. The more closely you look at this process, the denser the particles of the fluctuation become. The following figure is the energy equation of these interactions calculated to the lowest relevant order in perturbation theory.

\begin{equation}
\left( \frac{\delta m}{m} \right)_{\substack{\mathrm{lowest} \\ \mathrm{order}}} = \frac{3}{2 \pi} \frac{e^2}{\hbar c} \ln{\frac{\lambda_{\mathrm{max.}}}{\lambda_{\mathrm{min.}}}}
\end{equation}

In this analysis, the ratio of the mass energy of multiple interactions to the normal mass of electrons is 1/137 times the logarithm of the maximum scale divided by the minimum scale. This formula was derived by Weisskopf a while ago and it is an expression that we do not talk much about today. This is because, according to the new renormalization method it is possible to argue without entering into the electron behavior in such detail. If we now return to and trust this basic idea, we have to cut off at the minimum distance $\lambda_{\mathrm{min.}}$. Therefore, if you follow Sugawara's idea that you do not think outside of electrical and gravitational interactions, it is best to take this minimum distance as a measure of the gravitational radius of the particle. If, on top of that, all energies, and thus all masses, are due to interactions, then we cannot help concluding that the left side of the expression is equal to 1. On the right side there is a logarithm whose value is known. Therefore, we get an expression that determines the value of the fine structure constant. Of course, this type of equation gives only the order of magnitude of the fine structure constant correctly, but it takes into account only the lowest order of interactions. If we attempt to pursue this program with a policy of eliminating all the physical constants consistently and reducing this to a problem of pure number and initial condition, we have to figure out the interaction in question more closely and calculate it.

In connection with this question, I asked a question as to what other elementary particles are included. The answer was that we could describe all elementary particles in the same way. Electrons, mesons, and all other elementary particles are properly considered as fluctuations of the basic electromagnetic and gravitational fields. What about the neutrinos? I was concerned about neutrinos as I was keenly interested in recent experimental observations on the absorption of neutrinos by Reines and Cowan and the reasonable values of the cross sections they found experimentally. We are indeed deeply interested in the major role of neutrino in elementary particle theory. We have reached the conclusion that neutrinos are released during the decay of the pi and mu meson, and during beta decay.

Sugawara agrees with this conclusion, and he said, in his opinion he can choose between the following two possibilities. That is, one considers (i) the neutrino as another field like an electromagnetic field and a gravitational field of rest mass 0. In this case we will use all three of these fields to describe the form of all elementary particles; or (ii) derive the neutrino from the gravitational field, that is, the result of rewriting the gravitational field in spinor form. He said that it is a subtle matter for him to decide between these two possibilities if he tried to pursue this pure field theory program. I told him that it was an important problem to be studied by all means.

I did not see the young men running here and there. I did not see a large organization with a computing machine. Sugawara talked to young people, thought about such problems, seemed to be walking around in his spare time and did not seem to have a specific program. At the end, I felt that it is appropriate to ask how to explain the differences between elementary particles, weak interaction with electrons and mu mesons and nucleons and how to explain the strong interaction between pi mesons and V-mesons. He says that it is reasonable to think that this difference is related to the type of fluctuation in the field, the microscopic form of the forward and backward movement illustrated in space-time.

I said ``Is not your argument a little classical?'' He laughed and reminded me of Feynman's work on obtaining quantum theory from any classical theory by using Lagrange's method in conjunction with the sum over paths. This method was also studied in Japan, especially from the viewpoint of its logical foundations.

I came back from the arguments between two great men of the past feeling that there are many things that require research on the many problems that they pointed out. I also felt that they were of the same opinion on a broad principle only in the following sense, that they do not see the problem in mutually contradictory ways. Meson theory, supported by Saigo Takamori, has been recognized by Sugawara as a kind of provisional expression of a more perfect theory. Saigo Takamori, on the other hand, is somewhat pessimistic about Sugawara's direction, but nevertheless the final theory is simple and they agreed in expecting that it would be much closer to the beautiful form, harmonized perfectly, than any of the ones we have today. At the same time, we think that we must solve many other problems at the same time to solve the problem of elementary particles.

One person has the traditional energy and courage of Japan, the other one has a love of Japanese beauty and harmony. I feel that this country should make a great contribution to the solution of this basic problem. Thank you for your attention.

\end{document}